\documentclass[a4paper,11pt]{article}
\pdfoutput=1 

\usepackage{jcappub} 

\usepackage[T1]{fontenc} 

\newcommand{\lb}{\left(} 
\newcommand{\rb}{\right)}
\newcommand{\lsb}{\left[} 
\newcommand{\rsb}{\right]} 
\newcommand{\lcb}{\left\{} 
\newcommand{\rcb}{\right\}} 
\newcommand{\lan}{\left\langle} 
\newcommand{\ran}{\right\rangle}

\usepackage{soul}

\title{\boldmath Spectators no more! How even unimportant fields can ruin your Primordial Black Hole model.}

\author{A. Wilkins}
\author[a]{A. Cable}

\affiliation[a]{Department of Physics, \\ Imperial College London,\\London , UK}

\emailAdd{alwwilkins@gmail.com}
\emailAdd{archie.cable18@imperial.ac.uk}

\abstract{In this work we terminate inflation during a phase of Constant Roll by means of a waterfall field coupled to the inflaton and a spectator field. The presence of a spectator field means that inflation does not end at a single point, $\phi_e$, but instead has some uncertainty resulting in a stochastic end of inflation. We find that even modestly coupled spectator fields can drastically increase the abundance of Primordial Black Holes (PBHs) formed by many orders of magnitude. The power spectrum created by the inflaton can be as little as $10^{-4}$ during a phase of Ultra Slow-Roll and still form a cosmologically relevant number of PBHs. We conclude that the presence of spectator fields, which very generically will alter the end of inflation, is an effect that cannot be ignored in realistic models of PBH formation.}

\begin{document}
\maketitle
\flushbottom

\section{\label{sec:intro} Introduction}
\noindent In the modern, standard formulation of cosmic inflation \cite{Starobinsky1980,Sato1981,Guth1981,Linde1982,Albrecht1982,Linde1983} the rapid cosmological expansion is driven by a single scalar field called the inflaton, $\phi$. However the presence of other scalar fields during inflation -- e.g. light moduli fields \cite{Turok1988, Damour1996,Kachru2003} or the Higgs field -- is a possibility worth considering. If these extra scalar fields have enough energy they can directly affect the dynamics of the inflaton and one is then dealing with multi-field inflation -- see \cite{Pinol2021} for an example of how to deal with this in the stochastic inflation approach. In this work however we are interested in scalar fields that do not affect the inflationary dynamics which are aptly called $\emph{spectator}$ fields $\sigma$. Even without directly affecting inflation these fields can be of critical importance. \\

\noindent In the \emph{curvaton} scenario \cite{Linde1997,Moroi2001,Lyth2002,Moroi2002,Lyth2003,Lyth2005,Vennin2015a} the dominant contribution to the primordial curvature perturbation comes from the spectator field (hence the name) and the inflaton's contribution is sub-dominant. This is usually achieved by having the inflaton decay into radiation before the curvaton decays so that there is a period where the curvaton is the dominant contribution to the energy budget. In some cases this can even trigger a short second period of inflation -- see \cite{Vennin2015a} for a full breakdown of all the possible configurations. Since the $60$s and $70$s it has been known that large enough perturbations at small scales could form an appreciable number of Primordial Black Holes (PBHs) \cite{1967SvA....10..602Z, Hawking1971} which in turn could be a Dark Matter candidate \cite{Hawking1971, CHAPLINE1975}. In principle this might be achieved in a curvaton scenario although it is difficult to get a large enough enhancement on small scales -- typically the primordial power spectrum needs to be $\mathcal{P}_{\zeta} \sim 10^{-2}$ -- without ruining the constraints imposed on larger scales by the CMB \cite{Akrami2020}. Even outside of the curvaton scenario spectator fields could still form PBHs from field bubbles \cite{Maeso2022}. In this work however we will be extending the ideas of Nassiri-Rad \textit{et. al} \cite{Nassiri-Rad2022} which utilises a hybrid inflation + spectator setup to generate a \emph{stochastic} end of inflation. \\

\noindent Hybrid inflation \cite{Linde1991,linde_hybrid_infl_1994} ends inflation at a fixed point $\phi_e$ regardless of the dynamics of the inflaton. This is achieved by introducing a massive scalar field $\chi$ known as the \emph{waterfall} field. While the behaviour of the waterfall field itself can have important consequences for the formation of PBHs \cite{Garcia-Bellido1996,Clesse2015,Tada2023,Tada2023a} we will instead be interested in coupling it to our spectator field \cite{Lyth2005} so that inflation doesn't always end at the same value of $\phi_e$. In the original work by \cite{Lyth2005} this was utilised to realise a curvaton scenario; however we will instead be investigating how this can induce a \emph{stochastic} end of inflation. As the formation of PBHs is such a finely tuned problem -- see e.g \cite{Cole2023} for a recent investigation -- we suspect that even a small impact on the end of inflation might drastically impact the formation of PBHs. \\

\noindent In this work we will utilise the stochastic inflation  \cite{Starobinsky1988, Nambu1988,Nambu1989,Mollerach1991,Salopek1991,Habib1992, Linde1994,Starobinsky1994} formalism as it can accurately describe the evolution of non-linear perturbations. This is achieved by splitting the inflationary perturbations into short- and long-wavelength components where the long-wavelength perturbations can be treated as effectively classical. The initially short-wavelength quantum perturbations are stretched by the inflationary expansion and impact the dynamical equations by the inclusion of a classical random noise term which is a well established approximation for the infrared (IR) behaviour of quantum fields in inflationary spacetimes \cite{Tsamis2005,Finelli2009,Finelli2010,Garbrecht2014,Garbrecht2015,Moss2017} -- see \cite{Cable2021,Cable2022} however for how to correct the standard picture for massive test fields and see \cite{Cohen2021b} for next-to-next-to leading order corrections to the standard stochastic framework. Because the dynamics are now stochastic it means that the time taken (measured in e-folds) to reach $\phi_e$ is also a stochastic quantity, denoted by $\mathcal{N}$. The stochastic $\delta \mathcal{N}$ formalism \cite{Enqvist2008, Fujita2013, Fujita2014, Vennin2015} allows one to compute the coarse-grained comoving curvature perturbation on uniform energy density time-slices $\zeta_{cg}$ through:
\begin{eqnarray}
\mathcal{N} - \left\langle \mathcal{N}\right\rangle = \zeta_{cg} = \dfrac{1}{(2\pi)^{3/2}}\int_{k_{in}}^{k_{end}}\mathrm{d}\vec{k}\zeta_{\vec{k}}e^{i\vec{k}\cdot\vec{x}} \label{eq:Rcg_defn}
\end{eqnarray}  
which is the usual comoving curvature perturbation $\zeta$, coarse-grained between scales $k_{in}$, the scale that crossed the Hubble radius at initial time, and $k_{end}$, the scale that crosses out the Hubble radius at final time. Therefore to compute curvature perturbations all one needs to do is obtain the first-passage-time (FPT) probability distribution function (PDF) to reach the end of inflation $\rho (\mathcal{N})$. This formalism has been successfully utilised to study the formation of PBHs in many different situations -- see e.g. \cite{Pattison2017,Pattison2021,Rigopoulos2021,Animali2022} -- but it is generally difficult to accurately resolve the tail of the distribution\footnote{Which is typically the crucial part of the distribution for PBH formation.} in all but the simplest scenarios. A numerical method using importance sampling \cite{Jackson2022, Tomberg2022} can be fruitful although there is no general analytic procedure\footnote{While working on this paper Tomberg has expanded his numerical method outlined in \cite{Tomberg2022} to constant roll generally in a semi-analytic way \cite{Tomberg2023}. We emphasise that our approach is different in a couple of crucial ways. Firstly Tomberg does not utilise explicitly the H-J approach and secondly Tomberg analyses a period of CR after a period of USR with $\varepsilon_2 < 0$ i.e. with the inflaton \emph{speeding up} afterwards (this paper and \cite{Tomberg2023} use a different sign convention for $\varepsilon_2$.). Instead we are primarily interested in analysing when $\varepsilon_2 > 0$ i.e. when the inflaton $\emph{slows down}$ and this includes the USR case. We agree that investigating the effects of a CR phase with $\varepsilon_2 < 0$ after the USR phase is important and we leave the analysis of this in our approach to future work.} and it was shown in \cite{Rigopoulos2023} how the $\rho(\mathcal{N})$ can be obtained using Renormalisation Group techniques. \\

\noindent We begin in section \ref{sec:CR} by describing how a general period of Constant-Roll (CR) -- defined by a temporally-constant second slow-roll parameter $\varepsilon_2$ -- can be described using the Hamilton-Jacobi (H-J) formulation of Stochastic Inflation \cite{Salopek1990,Salopek1991}. We show how it can compute quantities such as the power spectrum (\ref{eq: classical power spectrum}) and spectral tilt (\ref{eq:classical ns}) in section \ref{sec:CR_observables} before treating the FPT problem in section \ref{sec:FPT_CR}. We defer technical computations of the FPT problem for a general linear $\phi$ dependence to Appendix \ref{sec:FPT Heat}. The advantage of the H-J approach is that the equation of motion for the inflaton is always one dimensional -- see (\ref{eq:H-J EOM}) -- even outside of an attractor phase like Slow-Roll (SR) making the equations much easier to solve. There have been some concerns surrounding the use of the H-J in such a non-attractor phase as some suggest that the stochastic fluctuations should perturb \emph{away} from the original H-J classical solution. It has been shown that classical perturbations quickly die away -- see e.g. Appendix A.3 of \cite{Wilkins2023} -- but the treatment involving quantum fluctuations is to appear in an upcoming publication \cite{Prokopec2023}. For now we point to the results of Tomberg \cite{Tomberg2022} who has shown numerically that even in USR, perturbations \emph{along} the classical trajectory reproduce the correct results and as the H-J trajectory \emph{is} the classical one we are justified in using this economical approach\footnote{See also the recent investigation into the behaviour of the noise functions \cite{Mishra2023} which also suggests that the noise points along the classical trajectory.}. \\

\noindent We then consider a concrete model to generate a \emph{stochastic end of inflation} in section \ref{sec:waterfall}. This is the waterfall mechanism of hybrid inflation achieved by coupling a heavy waterfall field $\chi$ to the inflaton and a spectator field $\sigma$. We find that the end of inflation $\phi_e$ obeys its own stochastic equation of motion (\ref{eq: phie EOM}) which must be combined with the equation of motion for the inflaton during CR (\ref{eq:CR_EOM}). Doing so results in modifying the noise term -- see (\ref{eq:tildephie_noise}) -- enhancing the stochastic effects. We defer the computation of the validity of the waterfall mechanism during CR generally to Appendix \ref{app:sudden-end} and find that the addition of a spectator field generically enhances the sudden-end approximation. \\

\noindent We then investigate the formation of PBHs with and without a spectator field in section \ref{sec:CR PBHs}. We investigate the validity of two commonly used approximations without a spectator; the Gaussian limit and an expansion in the tail. After confirming the validity of the tail expansion for sufficiently large $\varepsilon_2$ we use this to investigate the effect a spectator field has on PBH formation during USR. We find that the presence of a spectator can massively increase the abundance of PBHs; even a modestly coupled spectator can increase the abundance by 25 orders of magnitude. We then investigate the effect a spectator has on the power spectrum required to produce an appreciable number of PBHs for CR generally. Here we see that the inflationary power spectrum can be smaller than $10^{-3}$ and still violate current observational constraints on PBHs -- see e.g. \cite{Carr2020,Green2020} for a review of these. We summarise our thoughts and outline future research directions in section \ref{sec:conc}.

\section{\label{sec:CR}Constant-Roll in the Hamilton-Jacobi formalism}
\begin{figure}[t!]
\centering
    \includegraphics[width=.95\linewidth]{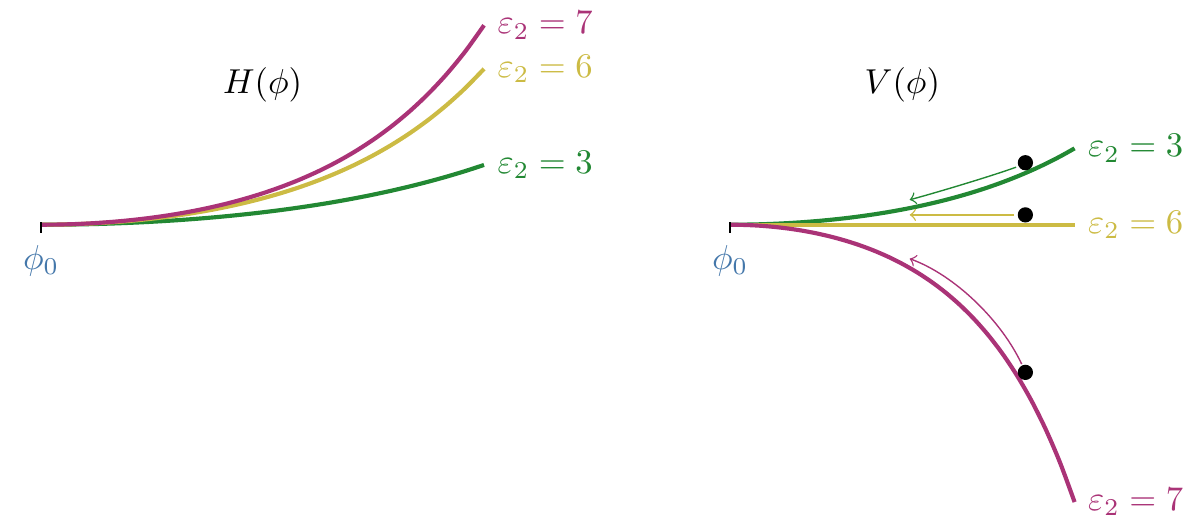}
    \caption{The Hubble parameter $H(\phi)$ (left) and potential $V(\phi)$ (right) in CR for different values of $\varepsilon_2$. In the right panel the arrows indicate the trajectories the inflaton would be taking classically during this period of CR. The potential heights have been altered by hand to coincide at $\phi_0$} 
    \label{fig:H and V}
\end{figure}

Throughout this paper we will be considering a period of Constant-Roll (CR) inflation as a testbed. To properly describe it we first introduce the definitions of the first two slow-roll parameters:
\begin{eqnarray}
    \varepsilon_1 \equiv -\dfrac{1}{H}\dfrac{\mathrm{d}H}{\mathrm{d}\alpha} \label{eq:epsilon1_defn} \\
     \varepsilon_2 \equiv -\dfrac{1}{\varepsilon_1}\dfrac{\mathrm{d}\varepsilon_1}{\mathrm{d}\alpha}, \label{eq:epsilon2_defn}
\end{eqnarray}
where $\alpha$ is the number of e-folds and $H$ is the Hubble expansion parameter. A period of CR is simply described by the fact that $\varepsilon_2$ is constant, $\partial_\alpha\varepsilon_2=0$. In the Hamilton-Jacobi (H-J) formulation of Stochastic Inflation \cite{Salopek1990,Salopek1991} $H$ is a function of the inflaton field only and has no explicit time dependence. It can be obtained by solving the Hamilton-Jacobi equation:
\begin{eqnarray}
\lb \dfrac{\mathrm{d}H}{\mathrm{d}\phi}\rb^2 = \dfrac{3}{2}H^2 - \dfrac{1}{2}V(\phi). \label{eq: H-J equation}
\end{eqnarray}
The stochastic equation of motion for the inflaton reads:
\begin{eqnarray}
\dfrac{\mathrm{d}\phi}{\mathrm{d}\alpha} = -2\dfrac{1}{H}\dfrac{\mathrm{d}H}{\mathrm{d}\phi} + A_{\phi}\xi(\alpha), \label{eq:H-J EOM} 
\end{eqnarray} 
where $A_{\phi}$ is the noise amplitude and $\xi$ is a Gaussian noise term with unit variance. For now we leave the noise amplitude general although it is common to take the de Sitter value $A_{\phi} = H/2\pi$. Utilising the fact that $\varepsilon_1 = 2(H_{,\phi}/H )^2$ and assuming that $\varepsilon_2 \neq 0$ we obtain:
\begin{eqnarray}
    \varepsilon_1 &=& \dfrac{\varepsilon_{2}^{2}}{8}\lb \phi - \phi_0\rb^2 \\
    H(\phi) &=& H_0\exp \lsb \dfrac{\varepsilon_{2}}{8}\lb \phi - \phi_0\rb^2\rsb \label{eq:H_CR}\\
    V(\phi) &=& H_{0}^2 \lb 3 - \dfrac{\varepsilon_2}{2}\rb \exp \lsb \dfrac{\varepsilon_{2}}{4}\lb \phi - \phi_0\rb^2\rsb,\label{eq:V_CR}
\end{eqnarray}
where $\phi_0$ is the value of the inflaton field when the classical drift is zero\footnote{When solving the H-J equation (\ref{eq: H-J equation}) it is necessary to introduce a constant of integration and we find expressing this in terms of $\phi_0$ to be the most convenient choice. This constant of integration is related to the momentum of the field at the start of the CR phase by $\Pi_{in} = -\varepsilon_2 (\phi_{in} -\phi_0)H_{in}/2$ where $\Pi \equiv H \partial_{\alpha}\phi$.} and $H_0$ is the value of the Hubble parameter at this point. We plot $H(\phi)$ and $V(\phi)$ for three different positive values of $\varepsilon_2$ in Fig.~\ref{fig:H and V}. Here we can see that while increasing the value of $\varepsilon_2$ steepens the curve for $H(\phi)$ it does not make much qualitative difference. However the difference is far more profound for the potential. If $\varepsilon_2 <0$ (which we generally ignore in this paper as they are not interesting for PBH formation) then the inflaton speeds up and $\varepsilon_1 \rightarrow 1$ -- see equation (\ref{eq: SR1_CR_alpha}). For $ 0 < \varepsilon_2 < 6$ the potential corresponds to a curve that gently flattens as it approaches $\phi_0$ with the field slowing down. However at the critical value of $\varepsilon_2 = 6$ -- commonly known as the Ultra Slow-Roll (USR) \cite{Tsamis2004, Kinney2005, Namjoo2013, Martin2013, Dimopoulos2017, Salvio2018, Pattison2018} regime\footnote{There are slightly contradictory definitions of a period of Ultra Slow-Roll in the literature. Sometimes it is defined by $\varepsilon_2 = \pm 6$, depending on whether there is a minus sign in the definition (\ref{eq:epsilon2_defn}). On other occasions it is defined by the presence of a region where $V_{,\phi} = 0$ in the potential, which is equivalent to $\eta \equiv -\varepsilon_1 + \varepsilon_2/2 = -3$. However as realistic inflationary potentials do not usually permit a region where $V_{,\phi} = 0$ exactly but instead where it is $V_{,\phi} \approx 0$ then both definitions are approximately true and the subtle differences aren't relevant.} -- the potential vanishes and we have the inflaton sliding in a de Sitter background. When $\varepsilon_2 > 6$ then the potential resembles the red curve in the right plot of Fig.~\ref{fig:H and V} and the inflaton is actually \emph{climbing} a slope and is therefore slowing down even faster than a period of USR. While it may seem difficult to realise an inflationary potential with this kind of feature -- see \cite{Briaud2023} for a recent investigation -- there have been many studies in the literature involving overcoming barriers \cite{Mishra2020,Zheng2021,Inomata2021,Cai2022,Cai2022a} with a focus on enhancing the curvature perturbation rapidly. It is therefore clear that varying the value of $\varepsilon_2$ appropriately can describe many different kinds of inflationary behaviour. 
\subsection{\label{sec:CR_observables}Observables from CR}
It was shown in \cite{Vennin2015} how cosmological observables could be computed in the stochastic inflationary formalism during a period of SR. This involved integrals that could not be solved analytically but fortunately they could be solved using a saddle-point approximation. It was shown that this saddle-point approximation was only valid when the classicality parameter $\eta_{cl} \ll 1$. As the name suggests the applicability of the saddle-point approximation corresponds to when one is in the \emph{semi-classical} regime, i.e. observables are closely described by the classical values obtained by e.g. the standard $\delta N$ formalism. It was shown in \cite{Rigopoulos2021} -- see also \cite{Wilkins2023} -- how these formulae could also be valid during a period of USR. Here we adapt the formulae for general use in CR. By performing this saddle point approximation we find that the \emph{semi-classical} formulae for the following quantities:\\
The average e-fold time, $\left\langle \mathcal{N}\right\rangle$, 
\begin{eqnarray}
\left\langle \mathcal{N}\right\rangle\vert_{cl} &=&  \int_{\phi_{e}}^{\phi}\dfrac{\mathrm{d}x}{M_{\mathrm{p}}}\dfrac{1}{\sqrt{2\varepsilon_1}(x)} =  \dfrac{2}{\varepsilon_2}\ln \lb \dfrac{\phi_{in} - \phi_0}{\phi_e - \phi_0}\rb .\label{eq: classical average e-fold} 
\end{eqnarray}
The deviation from the average e-fold time (or variance), $\delta \mathcal{N}^2 = \left\langle \mathcal{N}^2\right\rangle - \left\langle \mathcal{N}\right\rangle^2$,
\begin{eqnarray}
\delta \mathcal{N}^2\vert_{cl} &=& \dfrac{1}{4}\int_{\phi_{end}}^{\phi}\dfrac{\mathrm{d}x}{M_{\mathrm{p}}^4}\dfrac{H^{5}(x)}{H_{,x}^{3}(x)} \approx  \dfrac{2H_{0}^2}{\varepsilon_{1,in}\varepsilon_2}\lb e^{\varepsilon_2 N_{cl}} -1\rb .\label{eq: classical var e-fold} 
\end{eqnarray}
The power spectrum of curvature perturbations, $\mathcal{P}_{\zeta}\vert_{cl}$,
\begin{eqnarray}
\mathcal{P}_{\zeta}\vert_{cl} &=&  \dfrac{H^{2}(\phi)}{\varepsilon_{1}} \approx \dfrac{H_{0}^2}{\varepsilon_{1,in}}e^{\varepsilon_2 N_{cl}} .\label{eq: classical power spectrum}
\end{eqnarray}
The local non-Gaussianity, $f_{\scalebox{0.5}{$\mathrm{NL}$}} $,
\begin{eqnarray}
f_{\scalebox{0.5}{$\mathrm{NL}$}}\vert_{cl} &=& \dfrac{5}{24}\lsb 16\lb\dfrac{H_{,\phi}}{H}\rb^2 - 8\dfrac{H_{,\phi\phi}}{H}\rsb  = \dfrac{5}{24}\lb 4\varepsilon_{1,in}e^{-\varepsilon_2 N_{cl}} -2\varepsilon_2\rb .\label{eq:classical fNL}
\end{eqnarray}
The spectral tilt, $n_{\phi}$,
\begin{eqnarray}
n_{\phi} \vert_{cl} &=& 1-\lsb 8\lb\dfrac{H_{,\phi}}{H}\rb^2 - 4\dfrac{H_{,\phi\phi}}{H}\rsb = 1-2\varepsilon_{1,in}e^{-\varepsilon_2 N_{cl}} +\varepsilon_2 .\label{eq:classical ns} 
\end{eqnarray}
The classicality parameter, $\eta_{cl}$,
\begin{eqnarray}
\eta_{cl} &=& \left| \dfrac{3}{2}H^2 - \dfrac{H_{,\phi\phi}H^3}{2H_{,\phi}^{2}}\right| = \mathcal{P}_{\zeta}^2\vert_{cl} \left| \dfrac{7}{2}\varepsilon_{1,in}e^{-\varepsilon_2 N_{cl}} - \dfrac{\varepsilon_2}{4}\right| .\label{eq: classicality criterion}
\end{eqnarray}
The equation for the power spectrum (\ref{eq: classical power spectrum}) is particularly interesting as it explicitly shows what has been found elsewhere \cite{Byrnes2019,Carrilho2019} that the power spectrum grows quicker as $\varepsilon_2$ is increased. This is why it was originally thought \cite{Byrnes2019} that the steepest possible growth was given by $e^{6N_{cl}}$ as this corresponds to USR and a plateau in the potential. However, faster growth is possible if, as others have done  \cite{Mishra2020,Zheng2021,Inomata2021,Cai2022,Cai2022a}, you construct your potential such that there is a period where the field is climbing up and overcoming a barrier as shown by the red line in the right plot of Fig.~\ref{fig:H and V} -- see \cite{Tasinato2020} for an analytic insight into how this works. In multi-field scenarios this fast growth can also be achieved by fast turns in field space -- see e.g. \cite{Palma2020, Fumagalli2023}.\footnote{The authors thank the referee for bringing these results to our attention.} Equation (\ref{eq: classical power spectrum}) can also be straightforwardly inverted to inform how many e-folds of CR are required to reach a desired value of the power spectrum. 

\subsection{\label{sec:FPT_CR}Computing First-Passage-Time quantities during CR}
\begin{figure}[t!]
\centering
    \includegraphics[width=.8\linewidth]{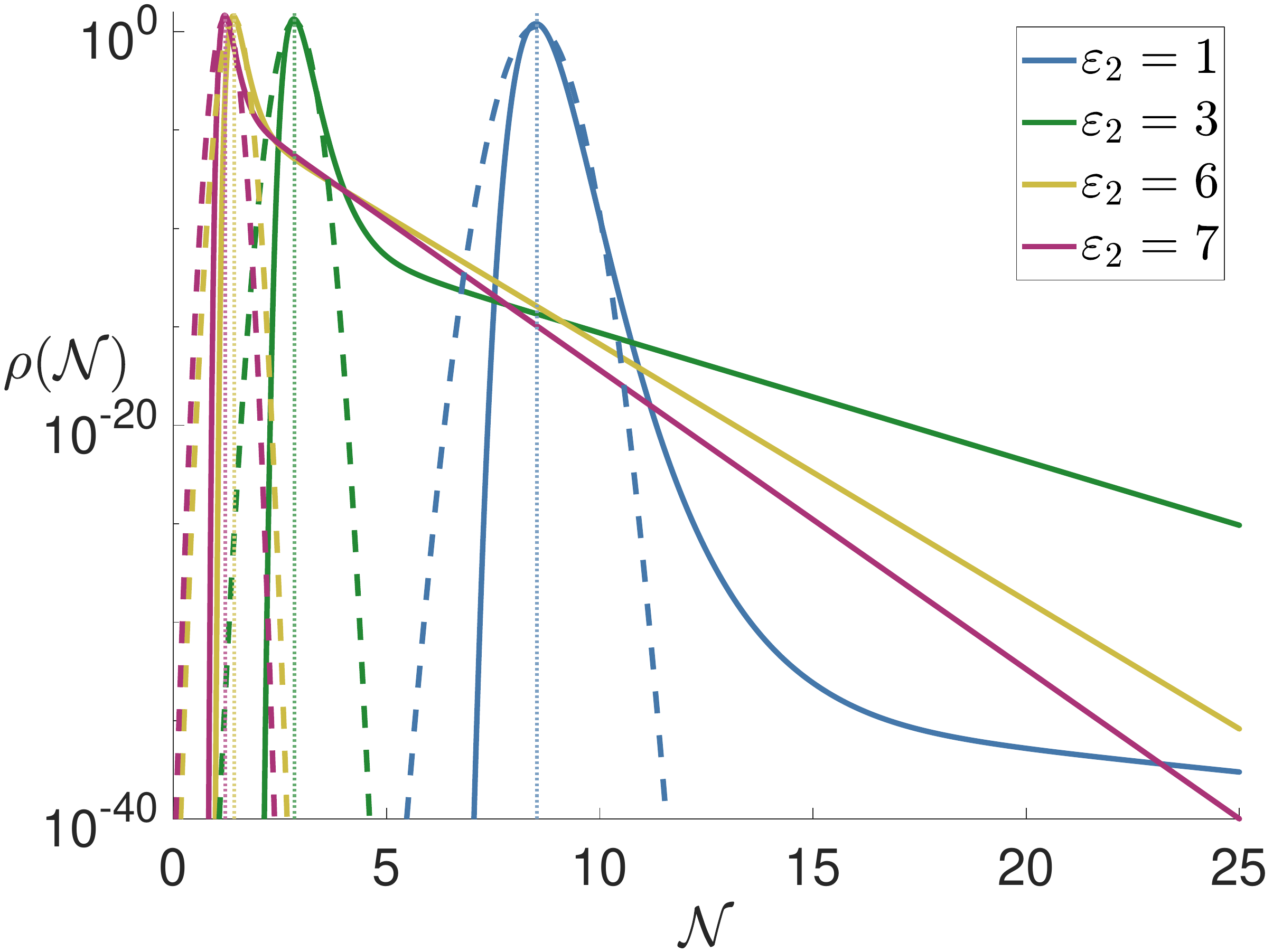}
    \caption{The PDF $\rho (\mathcal{N})$ for the inflaton to reach $\phi_e$ during CR for four different values of $\varepsilon_2$ shown log scale. The solid line corresponds to the full exact PDF (\ref{eq:rhoN CR}) whereas the dashed lines correspond to the NLO expansion around the classical limit (\ref{eq:rhoN_char_NLO}). In all cases the initial value of the first SR parameter $\varepsilon_{1,in} = 10^{-5}$, the scale of inflation is taken to be the maximum allowed by the CMB \cite{Akrami2020} $H_0 = 10^{-5}$ and the classical number of e-folds $N_{cl}$ realised during the CR phase is chosen so that the primordial power spectrum $\mathcal{P}_{\zeta} = 0.05$ -- see equation (\ref{eq: classical power spectrum}).} 
    \label{fig:rhoN_CR}
\end{figure}
Having determined the form of the $H(\phi)$ in (\ref{eq:H_CR}) we can then write the stochastic equation of motion (\ref{eq:H-J EOM}) during CR explicitly like so:
\begin{eqnarray}
    \dfrac{\mathrm{d}\phi}{\mathrm{d}\alpha} = -\dfrac{\varepsilon_2}{2}\lb \phi - \phi_0\rb + A_{\phi}\xi(\alpha).\label{eq:CR_EOM}
\end{eqnarray}
We can then ask ourselves, given that the inflaton was at some initial field value $\phi_{in}$ at time $\alpha_{in}$: how many e-folds $\mathcal{N}$ are required for it to reach $\phi_e$? As discussed around equation (\ref{eq:Rcg_defn}) the time taken is a stochastic quantity and the first-passage-time (FPT) PDF, $\rho (\mathcal{N})$ can be used to compute the coarse-grained curvature perturbation $\mathcal{\zeta}_{cg}$. In order to solve the FPT problem for the stochastic equation of motion (\ref{eq:CR_EOM}) we can modify the heat kernel techniques applied to a period of USR in \cite{Prokopec2018,Rigopoulos2021} -- see also appendix B of \cite{Wilkins2023} for more details -- to apply to CR more generally. The technical details of this have been deferred to appendix \ref{sec:FPT Heat} for which the FPT probability is given by (\ref{eq:rhoN HJ_append}):
\begin{eqnarray}
    \rho (\mathcal{N}) &=& \dfrac{\vert \varepsilon_2 \vert}{2\sqrt{\pi}}\exp \lsb -(n+1)\bar{U}^2\rsb \lsb n\sqrt{n+1}\Omega\mu -\sqrt{n}(n+1)\Omega \rsb \nonumber \\
&& -\dfrac{\vert \varepsilon_2 \vert}{2\sqrt{\pi}}\exp \lsb -n\bar{V}^2\rsb \lsb \sqrt{n}(n+1)\Omega-n\sqrt{n+1}\lb 2-e^{-\vert \varepsilon_2 \vert\Delta \mathcal{N}}\rb\Omega\mu \rsb e^{-Y} \nonumber \\
&& +\dfrac{\vert \varepsilon_2 \vert }{2}\Omega\mu\lsb \Omega\mu e^{-\vert \varepsilon_2 \vert\Delta\mathcal{N}}-n(2n+3)\Omega e^{-\vert \varepsilon_2 \vert\Delta\mathcal{N}/2}\rsb e^{-Y} \text{erfc}\lsb \sqrt{n}\bar{V}\rsb \label{eq:rhoN CR}
\end{eqnarray}
where
\begin{eqnarray}
Y &\equiv & \Omega^2 \mu \lb \dfrac{2n+1}{n+1}\mu +2ne^{-\vert \varepsilon_2 \vert\Delta \mathcal{N}/2} \rb\\
\bar{U} &\equiv & \Omega \lb \mu -  e^{-\vert \varepsilon_2 \vert\Delta \mathcal{N}/2}\rb\\
\bar{V} &\equiv & \Omega \lb 1 - \mu e^{-\vert \varepsilon_2 \vert\Delta \mathcal{N}/2} \rb\\
n &\equiv &\dfrac{1}{e^{\vert \varepsilon_2 \vert\Delta {\mathcal{N}}}-1}
\end{eqnarray}
and the parameters $\Omega$ and $\mu$ are given by\footnote{Note that the lack of a modulus in the exponent of (\ref{eq: mu defn CR}) is NOT a typo and in fact is the only difference between the $\varepsilon_2 <0 $ and $\varepsilon_2 > 0$ cases.}:
\begin{eqnarray}
    {\Omega} &\equiv  & S(\varepsilon_2)\dfrac{\sqrt{\vert \varepsilon_2 \vert}}{\sqrt{2}A_{{\phi}}}\lb {\phi}_{in} - {\phi}_0\rb = \dfrac{2\sqrt{\varepsilon_{1,in}}}{\sqrt{\vert \varepsilon_2 \vert}A_{{\phi}}} \label{eq: Omega defn CR} \\
{\mu} &\equiv &\dfrac{{\phi}_e - {\phi}_0 }{{\phi}_{in} - {\phi}_0} = e^{-\varepsilon_2 N_{cl}/2} \label{eq: mu defn CR}
\end{eqnarray}
where $\varepsilon_{1,in}$ is the value of the first SR parameter at the initial time $\alpha_{in}$ and $N_{cl}$ is the \emph{classical} number of e-folds it takes to traverse between $\phi_{in}$ and $\phi_e$. $S(x) = \pm 1$ for $x>0$ and $x < 0$ respectively. The $\varepsilon_2 = 0$ case will not be examined in detail in this work but a derivation of the FPT as well as PBH mass fraction can be found in Appendix \ref{sec:reallyCR}. An alternative method to compute the PDF of exit time, $\rho (\mathcal{N})$ using characteristic function techniques is outlined by Pattison et.al \cite{Pattison2017} -- see also section 4.3 of \cite{Wilkins2023} for what this looks like in the H-J formalism. We will focus on the expansion of the characteristic function around the classical limit. At leading order every trajectory takes the same amount of time and there are no coarse-grained comoving curvature perturbations. One must go to the NLO equation in \cite{Pattison2017} to obtain curvature perturbations with a Gaussian shape which for our CR system is:
\begin{eqnarray}
\rho_{\scalebox{0.5}{$\mathrm{NLO}$}}(\mathcal{N}) &=& \dfrac{1}{\sqrt{\pi \delta \mathcal{N}^2\vert_{cl}}}\exp \lsb -\lb\dfrac{\mathcal{N} -N_{cl}}{{\sqrt{\delta \mathcal{N}^2\vert_{cl}}}} \rb^2\rsb \label{eq:rhoN_char_NLO}
\end{eqnarray}
where $\delta \mathcal{N}^2\vert_{cl}$ is given by (\ref{eq: classical var e-fold}). We plot the PDF of exit time  $\rho (\mathcal{N})$ in Fig.~\ref{fig:rhoN_CR} using the full expression (\ref{eq:rhoN CR}) and the NLO expansion (\ref{eq:rhoN_char_NLO}) in solid and dashed lines respectively for four different values of $\varepsilon_2$. Each PDF corresponds to a scenario where the primordial power spectrum $\mathcal{P}_{\zeta} = 0.05$. We can see that while the NLO expansion is reasonable around the average value $N_{cl}$ it fails spectacularly at capturing the highly non-Gaussian tails as expected.

\section{\label{sec:waterfall} The waterfall mechanism during Constant-Roll}
During CR the equation of motion for the inflaton is given by (\ref{eq:CR_EOM}) and the first SR parameter grows or decays exponentially depending on the sign of $\varepsilon_2$:
\begin{eqnarray}
    \varepsilon_1 (\alpha) = \varepsilon_{1,in} \exp \lsb -\varepsilon_2 \lb \alpha - \alpha_{in}\rb\rsb \label{eq: SR1_CR_alpha}
\end{eqnarray}
If $\varepsilon_2 > 0$ then $\varepsilon_1 \rightarrow 0$ and so inflation \emph{cannot} end in the usual way during this phase (i.e. by $\varepsilon_1 \rightarrow 1$). Instead we must terminate inflation by some other mechanism before it reaches $\phi_0$. To do this we will employ a \emph{waterfall} mechanism.
\subsection{Terminating inflation at the same point}
\begin{figure}[t!]
\centering
    \includegraphics[width=.49\linewidth]{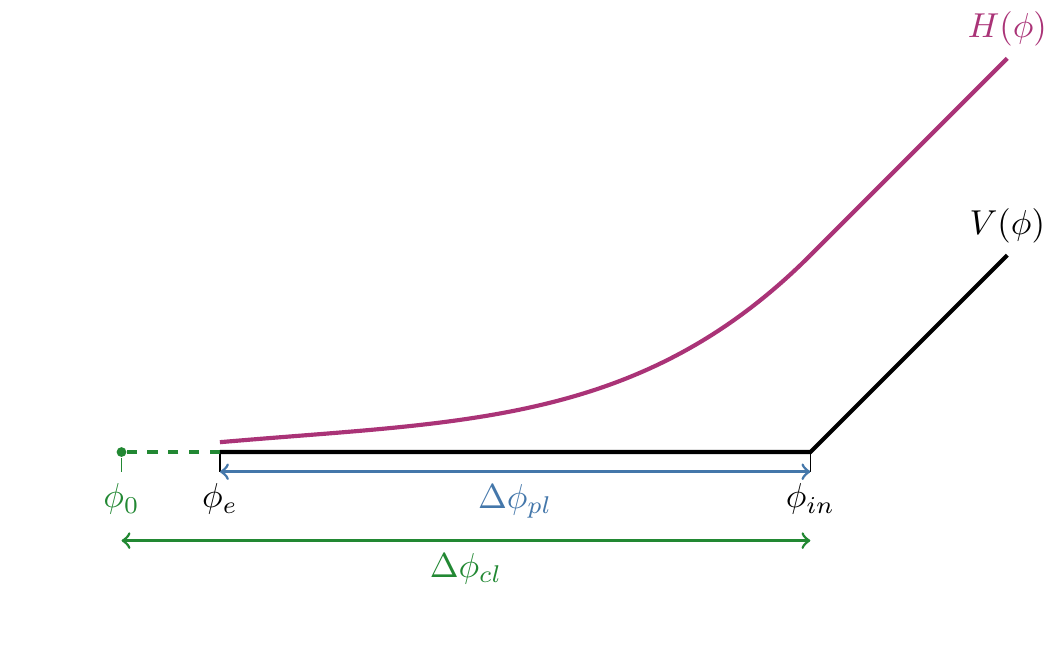}
    \includegraphics[width=.49\linewidth]{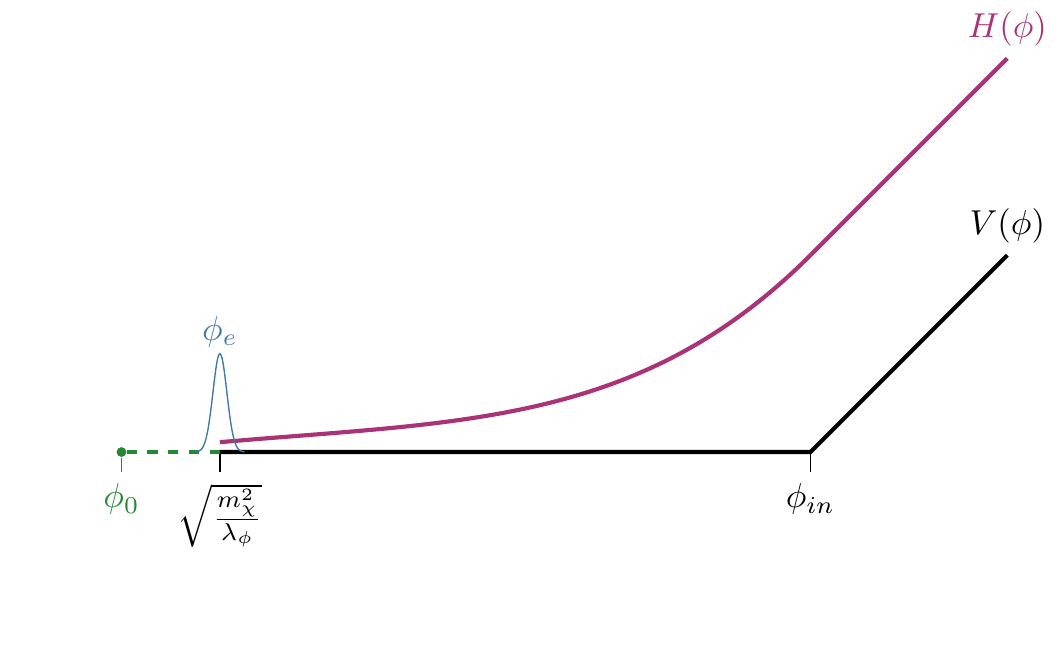}
    \caption{Terminating inflation on a plateau ($\varepsilon_2 = 6$) with a waterfall mechanism without a spectator field (left) and with a spectator field (right). The presence of a spectator field means that inflation doesn't always end at the same field value $\phi =m_{\chi}/ \sqrt{\lambda_{\phi}}$ but is instead a Gaussian distribution centred around this point.} 
    \label{fig:Plateau}
\end{figure}
The original hybrid inflation model \cite{Linde1991,linde_hybrid_infl_1994} introduces a massive scalar field $\chi$ known as the waterfall field. This waterfall field is coupled to the inflaton $\phi$ via the potential:
    \begin{eqnarray}
        W(\phi,\chi) = W_0 + V(\phi) - \frac{1}{2}m_{\chi}^2\chi^2 + \frac{1}{4}\lambda_{\chi}\chi^4 + \frac{1}{2}\lambda_{\phi}\phi^2\chi^2,
        \label{eq:waterfall_potential}
    \end{eqnarray}
where the mass of $\chi$ is given by $M(\phi)^2 = - m_{\chi}^2+\lambda_{\phi}\phi^2 $. In the early stages of inflation, $\phi$ is large so $M^2>0$ and the waterfall field will rapidly roll to its minimum at $\chi=0$. For most of inflation, this is where it will remain, having no effect on the inflaton. However, as the inflaton approaches $\phi_0$, there will be a point when $M^2$ becomes negative, causing $\chi$ to undergo a phase transition and triggering the end of inflation. We assume the sudden-end approximation such that this phase transition occurs very quickly, on a timescale far shorter than $H^{-1}$, which will be true provided $m_{\chi}^2\gg H^2$ \cite{linde_hybrid_infl_1994} -- see also Appendix \ref{app:sudden-end}. Therefore, the end of inflation is given by $\phi_e = m_{\chi}/\sqrt{\lambda_{\phi}}$. In the left plot of Fig.~\ref{fig:Plateau} we show what this would look like for a period of USR. We can see that inflation is terminated at the point $\phi_e > \phi_0$ when the first SR parameter is still much smaller than 1. This also means that the distance travelled on the plateau $\Delta \phi_{pl} \equiv \phi_{in} - \phi_e$ will be smaller than the distance that would have been travelled classically $\Delta \phi_{cl} \equiv \phi_{in} - \phi_0$. A more interesting setup is when inflation doesn't always end at the same point.
    
\subsection{Getting a stochastic end of inflation}
The waterfall mechanism can be adapted to include a spectator field $\sigma$ \cite{Lyth2005}:
    \begin{eqnarray}
        W(\phi,\chi,\sigma) = W_0 + V(\phi) - \frac{1}{2}m_{\chi}^2\chi^2 + \frac{1}{4}\lambda_{\chi}\chi^4 + \frac{1}{2}\lambda_{\phi}\phi^2\chi^2 - \frac{1}{2}\lambda_{\sigma}\sigma^2\chi^2 + \frac{1}{2}m_{\sigma}^2\sigma^2, \label{eq:waterfall+sigma_potential}
    \end{eqnarray}
where we have assumed that the spectator field evolves in a simple harmonic potential. Accounting for the spectator field, the mass of $\chi$ is now given by $M(\phi,\sigma)^2=-m_{\chi}^2+\lambda_{\phi}\phi^2-\lambda_{\sigma}\sigma^2$, where $\lambda_{\sigma,\phi}>0$. The addition of this spectator field speeds up the waterfall phase transition so we can once again take the sudden end approximation -- see Appendix \ref{app:sudden-end} for details. Now, inflation ends not at a fixed value of $\phi_e$ but on the \textit{surface}\footnote{Note in Lyth's original work the surface was an ellipse in phase space whereas here it is given by a hyperbola. Later we will see that this is necessary in our setup so as to avoid having an imaginary noise term in the equations of motion -- see (\ref{eq:phie noise}).} given by:
    \begin{eqnarray}
        \lambda_{\phi}\phi_{e}^2 - \lambda_{\sigma}\sigma^2 = m_{\chi}^2, \label{eq:end_inflation_surface}
    \end{eqnarray}
    although we still expect that on average inflation ends at the same point\footnote{Strictly speaking this is only true once the spectator field $\sigma$ has settled to its equilibrium distribution and $\lan \sigma \ran = 0$. Otherwise more generally $\lan \phi_e \ran$ will be time dependent and depend on the evolution of $\sigma$. We leave the study of the non-equilibrium case to future work.} i.e. $\lan \phi_e \ran = m_{\chi}/\sqrt{\lambda_{\phi}} \equiv \bar{\phi}_e $. This, combined with (\ref{eq:end_inflation_surface}) allows us to relate the connected two point function for the end of inflation with the spectator:
    \begin{eqnarray}
        \lan \phi_{e}^2\ran_C \equiv \lan \phi_{e}^2\ran - \lan \phi_e\ran^2 = \lan \phi_{e}^2\ran - \dfrac{m_{\chi}^2}{\lambda_{\phi}}= \frac{\lambda_{\sigma}}{\lambda_{\phi}} \lan \sigma^2\ran \equiv \tilde{\lambda} \lan \sigma^2\ran \label{eq:phie_sigma_2pt}
    \end{eqnarray}
We can see in the right plot of Fig.~\ref{fig:Plateau} that this means that inflation is no longer terminated at the same point every time but is instead given by a Gaussian distribution centred on $m_{\chi}/ \sqrt{\lambda_{\phi}}$ with variance given by (\ref{eq:phie_sigma_2pt}). In order to determine the precise value of this variance we examine the behaviour of the spectator field $\sigma$. Prior to the phase transition, the spectator field obeys the stochastic equations of motion:
\begin{eqnarray}
    \dfrac{\mathrm{d}\sigma}{\mathrm{d}\alpha}&=&\pi_{\sigma}+A_\sigma\xi(\alpha),\\
    \dfrac{\mathrm{d}\pi_\sigma}{\mathrm{d}\alpha}&=&-3\pi_\sigma-\frac{m_\sigma^2}{H^2}\sigma- \lb  3/2 - \nu \rb A_\sigma\xi(\alpha),\label{eq:full_sigma_eom}
\end{eqnarray}
where $\nu=\sqrt{\frac{9}{4}-\frac{m_\sigma^2}{H^2}}$ and the amplitude of the noise is given by \cite{Cable2021}:
    \begin{eqnarray}
        A_{\sigma}^2 = \frac{H^2}{4\pi^2}\frac{\Gamma (\nu)}{\Gamma (3/2)}\Gamma (5/2 -\nu). \label{eq:sigma_noise}
    \end{eqnarray}
Since the noise is one-dimensional for free fields, this can be recast as a one-dimensional stochastic equation:
    \begin{eqnarray}
        \frac{\mathrm{d}\sigma}{\mathrm{d}\alpha} = -(3/2-\nu)\sigma + A_{\sigma}\xi(\alpha).\label{eq:sigma EOM}
    \end{eqnarray}
    where the spectator field momentum is given by\footnote{An integration constant has been chosen such that $\pi_{\sigma} = 0$ at $\sigma = 0$.}:
    \begin{eqnarray}
        \pi_{\sigma} = -\lb 3/2 - \nu\rb\sigma. \label{eq:sigma_momentum}
    \end{eqnarray}
    Equations (\ref{eq:sigma_noise}), (\ref{eq:sigma EOM}) \& (\ref{eq:sigma_momentum}) therefore represent the generalisation of the standard spectator equations of motion now valid outside of SR and for $m_{\sigma} \lesssim H$ rather than the much stricter condition $m_{\sigma} \ll H$.
The two point function for $\sigma$ is then:
    \begin{eqnarray}
        \lan \sigma^2 (\alpha)\ran = \lsb \lan \sigma^2 (\alpha_{in})\ran - \frac{A_{\sigma}^2}{3-2\nu}\rsb \exp \lsb -\lb3-2\nu\rb\lb \alpha - \alpha_{in}\rb\rsb + \frac{A_{\sigma}^2}{3-2\nu}. \label{eq:sigma2ptfunc}
    \end{eqnarray}
    If we want the two point function for the end of inflation $\lan\phi_{e}^2\ran$ to appropriately match the spectator through (\ref{eq:phie_sigma_2pt}) we find that it should have the following equation of motion:
    \begin{eqnarray}
        \frac{\mathrm{d}\phi_e}{\mathrm{d}\alpha} = -\lb3/2-\nu\rb\lb \phi_e - \bar{\phi}_e\rb + A_{\phi_e}\xi(\alpha), \label{eq: phie EOM}
    \end{eqnarray}
    with the $\phi_e$ noise related to the $\sigma$ noise like so:
    \begin{eqnarray}
        A_{\phi_e}^2 = \tilde{\lambda}A_{\sigma}^2. \label{eq:phie noise}
    \end{eqnarray}
    If we make the transformation $\tilde{\phi} = \phi - \phi_e$ moving to a frame of reference where inflation always ends at the same field value, as \cite{Nassiri-Rad2022} do, we can combine the stochastic equations of motion for the inflaton during CR (\ref{eq:CR_EOM}) and for the end of inflation (\ref{eq: phie EOM}) like so:
\begin{eqnarray}
    \frac{\mathrm{d}\tilde{\phi}}{\mathrm{d}\alpha} = -\dfrac{\varepsilon_2}{2}\lb \tilde{\phi} - \tilde{\phi}_0 \rb + \lb3/2-\nu\rb\lb \phi_e - \bar{\phi}_e\rb + A_{{\phi}}\xi_1(\alpha) - A_{\phi_e}\xi_2(\alpha), \label{eq:tildephi EOM_2 noise}
\end{eqnarray}
where $\tilde{\phi}_0 = \phi_0 - \phi_e$ is now also a stochastic variable. This suggests that $\tilde{\phi}_0$ follows a Gaussian distribution centered at $\bar{\phi}_0 \equiv \phi_0 - \bar{\phi}_e$ with covariance $\langle \tilde{\phi}_{0}^2\rangle_C = \langle {\phi}_{e}^2\rangle_C$. These two moments for the $\tilde{\phi}_0$ distribution can be obtained by making the identification $\tilde{\phi}_0 = \bar{\phi}_0 + A_{\tilde{\phi}_0}\xi$ where
\begin{eqnarray}
    A_{\tilde{\phi}_0}^2 = \dfrac{A_{\phi_e}^2}{3-2\nu}.\label{eq:tildephi_0 noise}
\end{eqnarray}
This allows us to rewrite (\ref{eq:tildephi EOM_2 noise}) like so:
\begin{eqnarray}
    \frac{\mathrm{d}\tilde{\phi}}{\mathrm{d}\alpha} = -\dfrac{\varepsilon_2}{2}\lb \tilde{\phi} - \bar{\phi}_0 \rb + A_{{\phi}}\xi_1(\alpha) - A_{\phi_e}\xi_2(\alpha) + \dfrac{\varepsilon_2}{2}A_{\tilde{\phi}_0}\xi_3(\alpha), \label{eq:tildephi EOM_3 noise} 
\end{eqnarray}
where we have neglected the second gradient term in (\ref{eq:tildephi EOM_2 noise}) as it is generically very small compared to the first one\footnote{\label{footnoteR}To see this explicitly we find that the condition to neglect the second gradient term is simply $1 \gg \sqrt{3-2\nu}\sqrt{\mathcal{P}_{\zeta}}R/2$ which is trivially true in the massless limit $\nu\rightarrow3/2$ and even true in the massive limit $\nu \rightarrow 0$ at PBH formation provided that $R \leq 1$, which corresponds to $\lambda_\sigma\lesssim\lambda_\phi$.}. Combining the three independent stochastic processes allows us to write:
\begin{eqnarray}
    \frac{\mathrm{d}\tilde{\phi}}{\mathrm{d}\alpha} = -\dfrac{\varepsilon_2}{2}\lb \tilde{\phi} - \bar{\phi}_0 \rb  + A_{\tilde{\phi}}\xi(\alpha), \label{eq:tildephi EOM_1 noise} 
\end{eqnarray}
where the new noise term is given by:
\begin{eqnarray}
        A_{\tilde{\phi}}^2 = \frac{H^2}{4\pi^2}\lsb 1 + R\lb 1 + \dfrac{\varepsilon_{2}^2}{4}\dfrac{1}{3-2\nu}\rb\rsb, \label{eq:tildephie_noise}
    \end{eqnarray}
    where we have introduced the parameter $R$ to measure the enhancement of the noise term induced by the presence of a spectator field:
\begin{eqnarray}
        R = \dfrac{A_{\phi_e}^2}{A_{\phi}^2} = \tilde{\lambda}\frac{\Gamma (\nu)}{\Gamma (3/2)}\Gamma (5/2 -\nu) \label{eq: R defn}
\end{eqnarray}
so that $R = 0$ recovers the spectatorless case. Before we discuss the formation of PBHs let us examine how this stochastic end of inflation modifies our observables. 

\subsection{Cosmological observables with a stochastic end of inflation}
The original motivation by Lyth \cite{Lyth2005} to create a setup with a variable end of inflation was such that $\sigma$ would act as the \emph{curvaton} and be the dominant contribution to the primordial power spectrum. However our goal is to create a setup where a true spectator field, that has a negligible impact on the CMB, can greatly affect the production of PBHs. Therefore in addition to the condition that $R \leq 1$ -- see footnote \ref{footnoteR} -- we see what limits should be placed on the mass of the spectator field $m_{\sigma}$. To investigate this we relate the observables calculated earlier for $R = 0$, (\ref{eq: classical average e-fold}) -- (\ref{eq: classicality criterion}), to those for general $R$:
\begin{eqnarray}
\langle \tilde{\mathcal{N}}\rangle\vert_{cl} &=& \left\langle \mathcal{N}\right\rangle\vert_{cl}\label{eq: classical average e-fold_tilde} \\
\delta \tilde{\mathcal{N}}^2\vert_{cl} &=& \lsb 1 + R\lb 1 + \dfrac{\varepsilon_{2}^2}{4}\dfrac{1}{3-2\nu}\rb \rsb\delta \mathcal{N}^2\vert_{cl}  \label{eq: classical var e-fold_tilde} \\
\tilde{\mathcal{P}}_{\zeta}\vert_{cl} &=&  \lsb 1 + R\lb 1 + \dfrac{\varepsilon_{2}^2}{4}\dfrac{1}{3-2\nu}\rb \rsb \mathcal{P}_{\zeta}\vert_{cl}\label{eq: classical power spectrum_tilde} \\
\tilde{f}_{\scalebox{0.5}{$\mathrm{NL}$}}\vert_{cl} &=& f_{\scalebox{0.5}{$\mathrm{NL}$}}\vert_{cl}\label{eq:classical fNL_tilde}\\
\tilde{n}_{\phi} \vert_{cl} &=& n_{\phi} \vert_{cl} \label{eq:classical ns_tilde} \\
\tilde{\eta}_{cl} &=&  \lsb 1 + R\lb 1 + \dfrac{\varepsilon_{2}^2}{4}\dfrac{1}{3-2\nu}\rb \rsb\eta_{cl}.\label{eq: classicality criterion_tilde}
\end{eqnarray} 
The only observable of consequence that is affected is the power spectrum. To be clear $\tilde{\mathcal{P}}_{\zeta}\vert_{cl}$ is the value of the primordial power spectrum that would be measured e.g. in the CMB whereas $\mathcal{P}_{\zeta}\vert_{cl}$ is the contribution from the inflaton alone. It is clear from (\ref{eq: classical power spectrum_tilde}) that the massless limit $m_{\sigma} \rightarrow 0$ offers a massive enhancement of the power spectrum and that in general for $m_{\sigma} \ll H$ one is in danger of realising the curvaton scenario. For the rest of this paper we will therefore consider $m_{\sigma}^2/H^2 \sim 0.69$ (more precisely $\nu = 5/4$). To see how this avoids the curvaton scenario while still being interesting consider first the USR example $R = 1$ (corresponding to $\tilde{\lambda}\sim1$) and $\varepsilon_2 = 6$. Then we can see through (\ref{eq: classical power spectrum_tilde}) that this increases the power spectrum by a factor of 19 suggesting it might greatly impact PBH production. However now consider a period of CR compatible with the CMB. Then $-1 \ll \varepsilon_2 < 0$ in which case $\tilde{\mathcal{P}}_{\zeta}\vert_{cl} \approx  (1+R)\mathcal{P}_{\zeta}\vert_{cl}$ which only doubles the amplitude for the maximally allowed value of $R = 1$. This suggests that the contribution to the primordial power spectrum observed in the CMB from the spectator field can be at most comparable to the inflaton's and therefore we are not considering a curvaton scenario.

\section{\label{sec:CR PBHs}PBH formation during CR with a stochastic end of inflation}
We will use the Press-Schecter formalism for computing the mass fraction assuming a near monochromatic peak in the power spectrum. This will provide an \emph{underestimate} of the true abundance of PBHs compared to peaks theory -- see e.g. Appendix A of \cite{Kitajima2021}. The mass fraction of PBHs, $\beta$, can then be computed from the probability distribution function (PDF) of the coarse-grained\footnote{There are some subtleties involved with using the coarse-grained curvature perturbation rather than the standard comoving curvature perturbation. The main issue is that $\zeta_{cg}$ is actually typically coarse-grained at the scale of the end of inflation hypersurface according to (\ref{eq:Rcg_defn}) whereas the smoothing required for to accurately compute the abundance of PBHs is on the scale of the perturbation itself given by (\ref{eq:Rcg defn_PBH}) and the two are not necessarily equivalent. There have been some attempts to relate the coarse-grained comoving curvature perturbation directly to quantities like the density contrast \cite{Tada2022} however these have relied on equating quantities computed using different window functions and as discussed before \cite{Young2020} this eliminates any gain in accuracy this procedure would hope to achieve.} scalar curvature perturbation $\zeta_{cg}$:
\begin{eqnarray}
\beta (M) &=& 2 \int_{\zeta_{c}}^{\infty} P(\zeta_{cg})~\mathrm{d}\zeta_{cg}, \label{eq:massfracdef} \\
\zeta_{cg}(\textbf{x}) &\equiv & (2\pi)^{-3/2} \int_{k > aH_{form}}\mathrm{d}\textbf{k}\zeta_{\textbf{k}}e^{i\textbf{k}\cdot\textbf{x}}, \label{eq:Rcg defn_PBH}
\end{eqnarray}
so that the mass fraction $\beta$ represents the area under the curve\footnote{Multiplied by a factor of 2 to account for the under-counting in Press-Schecter theory \cite{Press1974}.} of the PDF above some critical value, $\zeta_c$. Recall from (\ref{eq:Rcg_defn}) that the coarse-grained curvature perturbation is given by $\zeta_{cg}(\textbf{x}) = \mathcal{N} - \lan \mathcal{N}\ran$ where each ``point $x$'' is a Hubble-sized\footnote{Strictly speaking a coarse-grained sized patch will be larger than a Hubble-sized patch by a factor of $1/\nu^3$.} patch whose field value is represented by one of the trajectories in the random walk of the stochastic inflation equation (\ref{eq:H-J EOM}). The constraints on the mass fraction $\beta$ depend on the full post-inflationary evolution of the universe and the mass of the PBHs but for our simple purposes we will say that the strongest constraints correspond to $\beta < 10^{-24}$ and the weakest to $\beta < 10^{-2}$ -- see e.g. section 5.2 of \cite{Wilkins2023} for an overview of where these come from. 

Using the curvature perturbation, $\zeta$, to compute the PBH mass fraction can be dangerous \cite{Musco2019, Young2019, Germani2020, Young2020, Biagetti2021, Gow2021}. It is clear that to get the most accurate result one should instead replace $\zeta$ in (\ref{eq:massfracdef}) with the density contrast $\delta$ which is related to $\zeta$ in a highly non-linear way -- see \cite{Gow2022,Ferrante2023} for the most up to date approach for how to do this in general. In this work we are interested in the modification that the spectator field has on the PBH abundance so we are justified in the use of the simple expression (\ref{eq:massfracdef}) to compare the abundances with and without a spectator field. 
\subsection{\label{sec:PBH_nospec}Abundance without a spectator}
\begin{figure}[t!]
\centering
    \includegraphics[width=.8\linewidth]{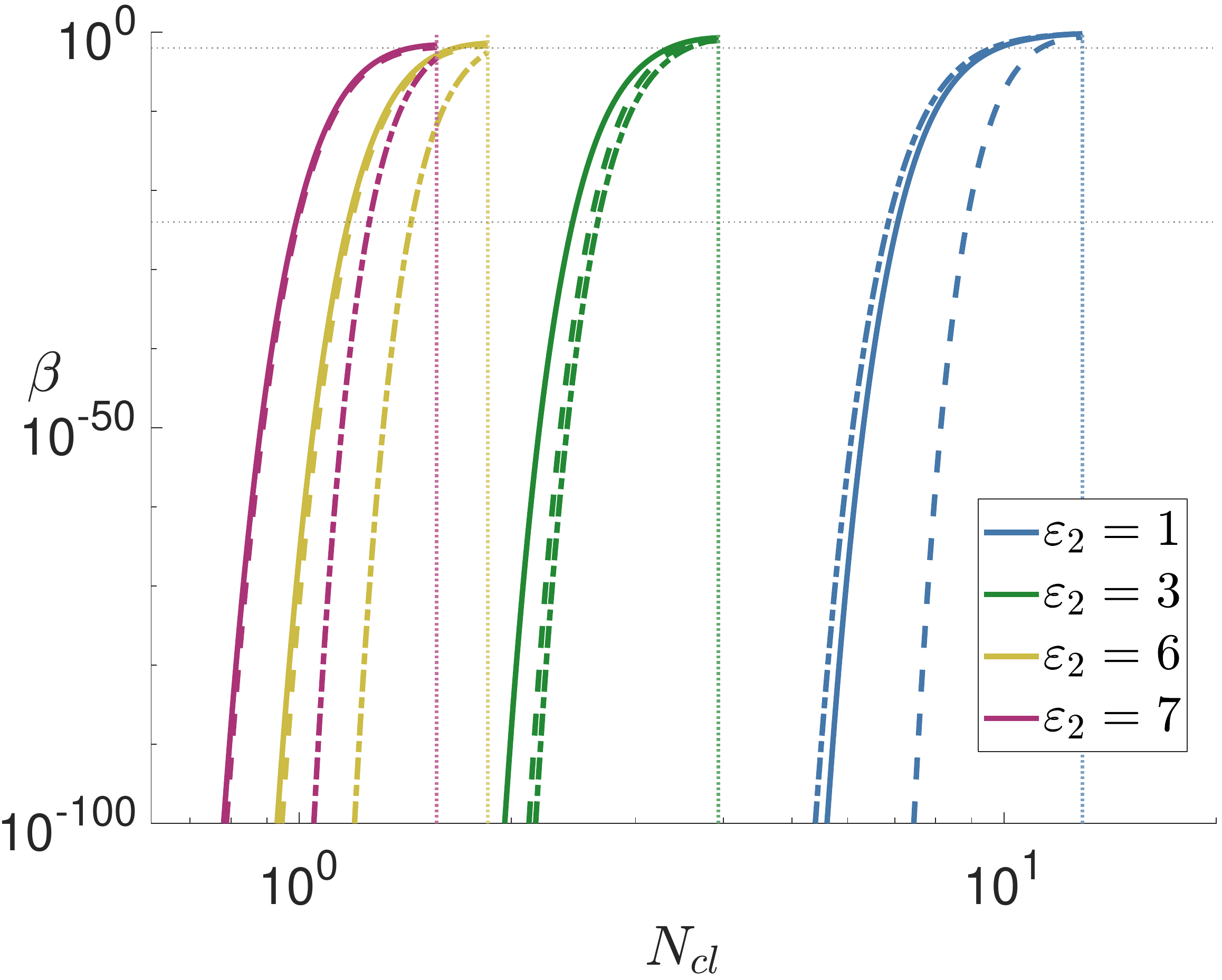}
     \caption{The mass fraction as a function of the classical number of e-folds spent in the CR phase $N_{cl}$ for different values of $\varepsilon_2$. The dashed and dot-dashed curves correspond the mass fraction as computed by an expansion around the tail (\ref{eq: Beta tail}) and an expansion around the peak (\ref{eq:Beta NLO}) of the PDF for exit times, while the solid curves represent the full expression (\ref{eq:massfrac_exactFULL_CR}) for the mass fraction. The horizontal dotted lines represent the strongest ($10^{-24}$) and weakest ($10^{-2}$) constraints on the abundance of PBHs. Any values in the parameter space that exceed these constraints are ruled out. The vertical dashed lines corresponds to the value ${\Omega}{\mu} = 1/\sqrt{2}$ which signifies entering the quantum diffusion regime and is the limit of validity for $N_{cl} \approx \lan \mathcal{N} \ran$. In all cases the initial value of the first SR parameter $\varepsilon_{1,in} = 10^{-5}$ and the scale of inflation is taken to be the maximum allowed by the CMB \cite{Akrami2020} $H_0 = 10^{-5}$.} 
     \label{fig: Beta_CR}
\end{figure}

Let us begin by examining how to compute the abundance of PBHs on a plateau where inflation always ends at the same point $\phi_e = m_{\chi}/ \sqrt{\lambda_{\phi}}$. This problem was solved in \cite{Rigopoulos2021} for the USR case and we generalise in Appendix \ref{sec:FPT linear} to the CR case with the result being given by (\ref{eq:massfrac_exactFULL_linear}):
\begin{eqnarray}
\beta &=& \dfrac{2\lsb e^{-{\Omega}^2{\mu}^2}-\text{erfc}({\Omega}{\mu}) - e^{Y_{c}}\text{erfc}(\sqrt{n_c}\bar{V}_c) + \text{erfc}(\sqrt{n_c + 1}\bar{U}_c) \rsb}{1 + \text{erf}({\Omega}{\mu}) + e^{-{\Omega}^2{\mu}^2}},\label{eq:massfrac_exactFULL_CR}\\
Y_c &\equiv & -{\Omega}^2{\mu}^2 \lsb 1 + e^{-\vert \varepsilon_2 \vert(\zeta_c + N_{cl})} +\dfrac{2n_c}{{\mu}} e^{-\vert \varepsilon_2 \vert(\zeta_c + N_{cl})/2} \rsb, \label{eq: Y defn_CR} \\
\bar{U}_c &\equiv & {\Omega}\lsb {\mu} -  e^{-\vert \varepsilon_2 \vert(\zeta_c + N_{cl})/2} \rsb,\label{eq: U defn_CR}\\
\bar{V}_c &\equiv & {\Omega}\lsb 1 - {\mu} e^{-\vert \varepsilon_2 \vert(\zeta_c + N_{cl})/2} \rsb,\label{eq: V defn_CR} \\
n_{c} &\equiv &\dfrac{1}{e^{\vert \varepsilon_2 \vert(\zeta_c + N_{cl})} -1}, \label{eq:nc defn_CR} 
\end{eqnarray}
where $\zeta_c$ is the cutoff for PBH formation which we take to be $1$ and we have approximated the average number of e-folds spent in the CR phase with the classical value $\lan \mathcal{N}\ran \approx N_{cl}$ which is valid for $\eta_{cl} \ll 1$. It was shown in \cite{Rigopoulos2021} that the inflaton is never dominated by quantum diffusion so this approximation is always valid. Erf and erfc are the standard error and complementary error functions respectively. If one expands the PDF of exit time $\rho (\mathcal{N})$ (\ref{eq:rhoN CR}) in the tail of the distribution, then the mass fraction (\ref{eq:massfrac_exactFULL_CR}) can be approximated by:
\begin{eqnarray}
\beta_{\scalebox{0.5}{Tail}} &\simeq &\dfrac{4}{\sqrt{\pi}}\Omega \mu ~e^{-\Omega^{2} \mu^{2}}~e^{-\varepsilon_2\zeta_c /2}, \label{eq: Beta tail} 
\end{eqnarray}
so that the mass fraction depends only on the combination $\Omega\mu$. On the other hand one can approximate the mass fraction based on the NLO approximation (\ref{eq:rhoN_char_NLO}) in which case the mass fraction is:
\begin{eqnarray}
    \beta_{\scalebox{0.5}{$\mathrm{NLO}$}} = \text{erfc}\lb \dfrac{\zeta_c}{\sqrt{\delta \mathcal{N}^2\vert_{cl}}}\rb \label{eq:Beta NLO}
\end{eqnarray}
These different approximations focus on the behaviour of very rare and very frequent fluctuations respectively. In Fig.~\ref{fig: Beta_CR} we examine the abundance of PBHs during CR for four different values of $\varepsilon_2$ as a function of the classical number of e-folds $N_{cl}$ spent in the CR phase with the solid curves showing the exact abundance (\ref{eq:massfrac_exactFULL_CR}). Unsurprisingly one needs a much longer CR phase for smaller values of $\varepsilon_2$ as perturbations take longer to grow. We compare these curves to the mass fraction from the tail (dashed lines) and from the NLO expansion (dot-dashed lines). We can see that for large values of $\varepsilon_2$ the tail expression (\ref{eq: Beta tail}) captures the full mass fraction very well whereas the NLO expansion (\ref{eq:Beta NLO}) is very poor, severely underestimating the abundance of PBHs. On the other hand at around $\varepsilon_2 = 3$ we can see that the two approximations are both almost equally poor, underestimating the abundance by many orders of magnitude. This suggests that the important region in the PDF for PBH formation in this case is neither in the tail or near the peak but somewhere in between. Finally for $\varepsilon_2 = 1$ we can see that the tail expression is very poor, severely underestimating the abundance of PBHs. On the other hand the NLO approximation matches the true value much better but now \emph{overpredicts} the abundance of PBHs. This emphasises that even in situations where one might believe it safe to do a NLO approximation -- as the tail of the distribution is not the important region for PBH formation -- one should still use the full expression.  

\subsection{Abundance with a spectator}
\begin{figure}[t!]
    \includegraphics[width=.49\linewidth]{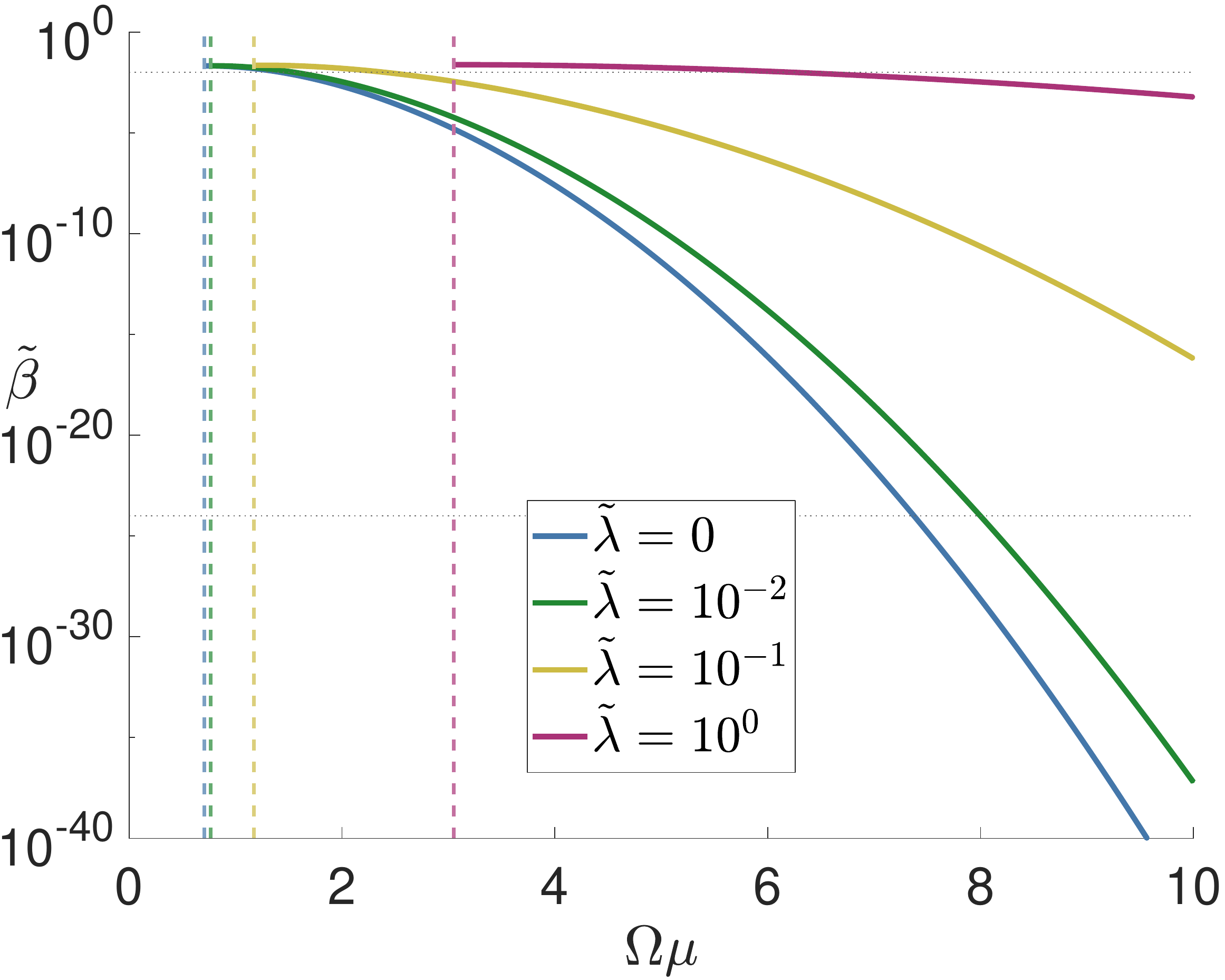}
    \includegraphics[width=.49\linewidth]{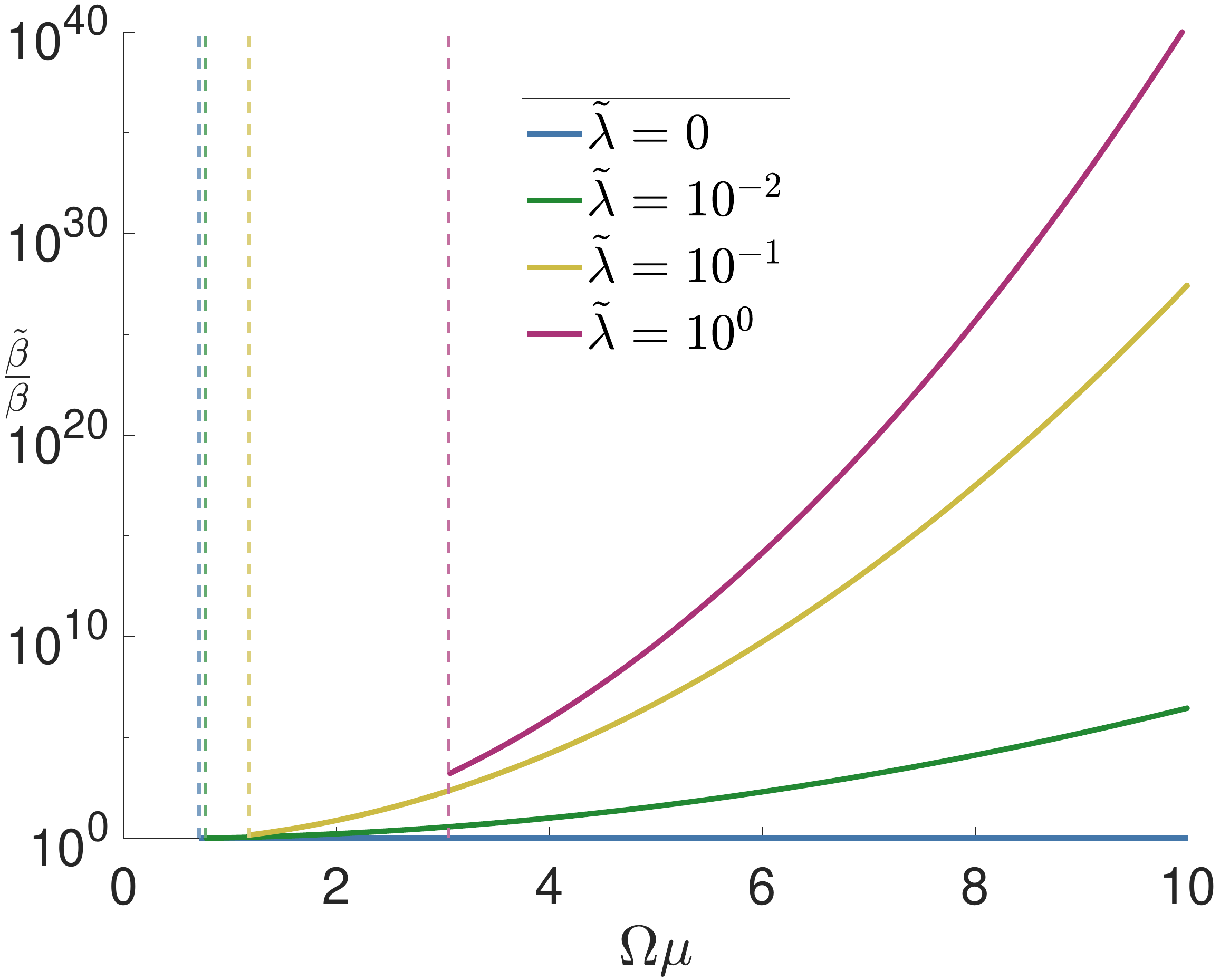}
     \caption{The mass fraction as a function of $\Omega\mu$ for different values of $R$ (left) and the enhancement (right) in the USR case $\varepsilon_2 = 6$ using the tail approximation (\ref{eq: Beta tail}) for $\nu = 5/4$. The horizontal dotted lines in the left plot represents the strongest ($10^{-24}$) and weakest ($10^{-2}$) constraints on the abundance of PBHs. The vertical dashed lines in both plots corresponds to the value $\tilde{\Omega}{\mu} = 1/\sqrt{2}$ which corresponds to entering the quantum diffusion regime and is the limit of validity for equation (\ref{eq: Beta tilde R}).} 
     \label{fig: beta_plat+spectator}
\end{figure}
The mass fraction without a spectator given by (\ref{eq:massfrac_exactFULL_CR}) is computed assuming that (\ref{eq:CR_EOM}) is the appropriate equation of motion. Here, to incorporate the effect of a stochastic boundary, we will use (\ref{eq:tildephi EOM_1 noise}) to compute the mass fraction which is of the same form as (\ref{eq:CR_EOM}) but with a different noise amplitude. The PDF for exit time and the mass fraction of PBHs will have the same form as they do in (\ref{eq:massfrac_exactFULL_CR}) \& (\ref{eq:massfrac_exactFULL_CR}) respectively under the replacement\footnote{In principle $\mu \rightarrow \tilde{\mu}$ but the two definitions are the same regardless of any modification to the noise so we do not do so for simplicity.} $\Omega \rightarrow \tilde{\Omega}$ which is defined like so:
\begin{eqnarray}
    \tilde{\Omega} &\equiv  & S(\varepsilon_2)\dfrac{\sqrt{\vert \varepsilon_2 \vert}}{\sqrt{2}A_{\tilde{\phi}}}\lb {\phi}_{in} - {\phi}_0\rb = \Omega 
 \dfrac{A_{\phi}}{A_{\tilde{\phi}}} = \Omega \dfrac{1}{\sqrt{1 + R\lb 1 + \varepsilon^2 /4(3-2\nu)\rb}}\label{eq:tildeOmega defn CR}
\end{eqnarray}
For USR where the tail approximation for the mass fraction (\ref{eq: Beta tail}) is good, we can simply write $\tilde{\beta}_{\scalebox{0.5}{Tail}}$ in terms of the old mass fraction, $\beta_{\scalebox{0.5}{Tail}}$, like so:
\begin{eqnarray}
        \tilde{\beta}_{\scalebox{0.5}{Tail}} = \beta_{\scalebox{0.5}{Tail}} \frac{1}{\sqrt{1 + R_{6}}}\exp \lsb \Omega^2\mu^2 \frac{R_{6}}{1+R_{6}}\rsb \label{eq: Beta tilde R},
    \end{eqnarray}
where $R_{\varepsilon_2} = R\lb1 + \frac{\varepsilon_{2}^2}{4}\frac{1}{(3-2\nu)}\rb$, as seen in (\ref{eq:tildephie_noise}). In Fig.~\ref{fig: beta_plat+spectator} we plot (\ref{eq: Beta tilde R}) for four values of $\tilde{\lambda}$ (left) and we also plot the enhancement, $\tilde{\beta}/\beta$, (right) for $\nu = 5/4$. We can clearly see that even small values of $\tilde{\lambda}$ can dramatically enhance the abundance of PBHs. For $\tilde{\lambda} = 0.1$ for instance -- which corresponds to the noise due to the stochastic boundary being roughly equal to the inflaton noise -- we can see that the mass fraction is enhanced by over 25 orders of magnitude for $\Omega\mu = 10$. This means that values in the parameter space that were previously well within observational constraints before the introduction of a spectator field can become ruled out with a spectator that is coupled to the waterfall field with a comparable strength as the inflaton. 

\begin{figure}[t!]
\centering
    \includegraphics[width=.8\linewidth]{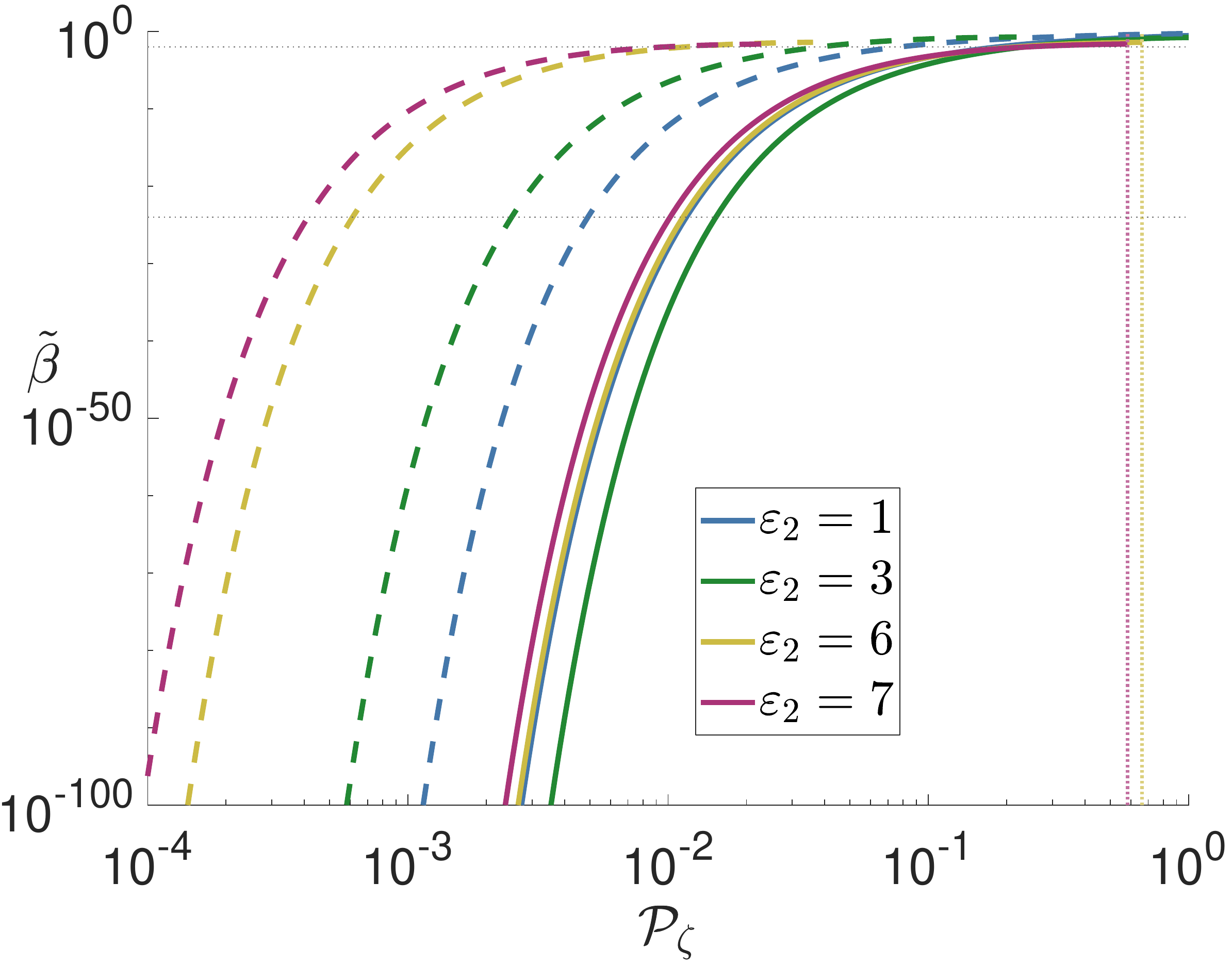}
    \caption{The mass fraction of PBHs -- computed from (\ref{eq:massfrac_exactFULL_CR}) -- as a function of the power spectrum $\mathcal{P}_{\zeta}$ -- from (\ref{eq: classical power spectrum}) -- generated by the inflaton without a spectator (solid curves) and for $\tilde{\lambda} = 1$, $\nu = 5/4$ (dashed curves) for different values of $\varepsilon_2$. The solid blue curve is hidden by the solid yellow curve. The horizontal dotted lines represent the strongest and weakest constraints on the abundance of PBHs. The vertical dashed lines correspond to entering the quantum diffusion regime and is the limit of validity for $N_{cl} \approx \lan \mathcal{N}\ran$.} 
    \label{fig:Beta_CR_Bigcompare}
\end{figure}

It is clear that the presence of a spectator field makes it \emph{easier} to form PBHs. To see this we consider how the abundance of PBHs depends on the primordial power spectrum $\mathcal{P}_{\zeta}$ generated by the inflaton. In Fig.~\ref{fig:Beta_CR_Bigcompare} we plot this for $\tilde{\lambda} = 0$ and $\tilde{\lambda} = 1$ (solid and dashed curves respectively) again for $\nu = 5/4$. Focusing firstly on the case without a spectator ($\tilde{\lambda} = 0$) we can see that increasing $\varepsilon_2$ from $3 \rightarrow 7$ means that more PBHs are formed at the same value of the power spectrum. However $\varepsilon_2 = 1$ and $\varepsilon_2 = 6$ have an almost identical dependence on the power spectrum suggesting that the relationship between $\varepsilon_2$ and the power spectrum required to form a given number of PBHs is non-trivial. If we now examine the $\tilde{\lambda} = 1$ case -- which corresponds to the inflaton and the spectator being coupled to the waterfall field with equivalent strength -- we see that the effect on the power spectrum is dramatic. For $\varepsilon_2 = 6$ we can see that the required value of the power spectrum drops by well over an order of magnitude. This reduction in value of the inflaton power spectrum needed to form PBHs can be mostly explained by the enhancement in the \emph{observed} power spectrum -- see equation (\ref{eq: classical power spectrum_tilde}) -- however not all of it can. We can say with confidence therefore that including a spectator field in this way can drastically alter the abundance of PBHs.

\section{\label{sec:conc}Discussion}
\noindent In this work we have investigated a period of Constant-Roll (CR) inflation -- using the Hamilton-Jacobi (H-J) formulation of Stochastic Inflation \cite{Salopek1990,Salopek1991} -- in the presence of two extra scalar fields, a waterfall field $\chi$ and a spectator field $\sigma$. We explicitly demonstrated a mechanism to obtain a stochastic equation of motion for the end of inflation itself (\ref{eq: phie EOM}). This equation was combined with the equation of motion for the inflaton during CR (\ref{eq:CR_EOM}) resulting in the conclusion that a modified end of inflation will \emph{enhance} the amplitude of the stochastic noise term (\ref{eq:tildephie_noise}). This enhancement depends on the value of the second SR parameter, $\varepsilon_2$, the mass of the spectator field $m_{\sigma}$ and the relative strength $\tilde{\lambda} $ of the couplings between $\phi$ and $\chi$ and between $\sigma$ and $\chi$. We highlighted how if the mass of the spectator field is too small then one is generically in the curvaton scenario and therefore chose a value of $m_{\sigma}^2 \sim 0.69$ to avoid this.\\

\noindent We examined the impact that this stochastic end of inflation has on the formation of PBHs during CR. We found that for even very modest values of the relative couplings, $\tilde{\lambda} = 0.1$, the enhancement the formation of PBHs during a period of USR could be 25 orders of magnitude -- see Fig.~\ref{fig: beta_plat+spectator}. This means that inflationary scenarios previously deemed ``safe'' (i.e. did not violate observational constraints) would now do so. We also investigated the dependence on the inflationary power spectrum required to form PBHs for several values of $\varepsilon_2$ -- see Fig.~\ref{fig:Beta_CR_Bigcompare}. Again the impact of the spectator is dramatic, reducing the amplitude of the power spectrum required to form PBHs that violate constraints to well below $10^{-3}$ for a period of USR. Even outside of the waterfall mechanism discussed here the presence of additional spectator fields during inflation will result in a modified end of inflation \cite{Assadullahi2016} meaning that the end of inflation $\phi_e$ will generically obey some sort of stochastic differential equation similar to (\ref{eq: phie EOM}). Therefore when the curvature perturbation, $\zeta$, is computed in the shifted frame, $\tilde{\phi} = \phi - \phi_e$, it will always be enhanced due to the effective amplification of quantum fluctuations through a modified noise term similar to (\ref{eq:tildephie_noise}). We therefore conclude that the presence of additional spectator fields during inflation will affect the formation of PBHs and, as explicitly shown here, can affect the formation so dramatically that their presence cannot be ignored a priori.\\

\noindent The CR + waterfall + spectator model we have discussed here can only form very light PBHs immediately after the end of inflation, which means they would not have lasted long enough to be dark matter. Our goal in this work was to merely highlight the dramatic impact a modified end of inflation can have on PBH abundance. Future work will address the problem of e.g. a USR phase followed by a CR phase with $\varepsilon_2 < 0$ as is more realistic. We would still expect a stochastic end of inflation to have a noticeable impact on the formation of PBHs in this case. To see this notice that the modification to the noise for this second CR phase would still be given by equation (\ref{eq:tildephie_noise}) which for a small negative value of $\varepsilon_2$ simplifies to $(1+R)H^2/4\pi^2$ suggesting that there may still be an appreciable modification to heavier PBHs outside of a curvaton scenario. Further, we have considered an explicit model where the end of inflation is no longer constant but instead follows a Gaussian distribution. It would be interesting to consider more generally how a stochastic end of inflation affects the formation of PBHs via a model-independent approach. We leave this to future work.

\acknowledgments
The authors would like to thank UKCosmo for providing a stimulating environment for discussions about this work and Christian Byrnes, Joe Jackson and David Wands in particular for their questions and insight. The authors would also like to thank the anonymous referee for their comments which have improved the quality of this work. 

\appendix
\section{\label{sec:FPT Heat}Computing First-Passage Times using Heat Kernel techniques}
In this appendix we will derive the First-Passage-Time (FPT) probabilities for some cosmologically relevant stochastic processes. In particular we will be exploiting the fact that the probability distribution $\rho (\mathcal{N})$ for number of e-folds to reach a field value $\phi_e$ can be obtained from the knowledge of the probability $P(\phi,\alpha)$ that enters the corresponding Fokker-Planck (F-P) equation. These are related through \cite{VanKampen2007}:
\begin{eqnarray}
\int_{\mathcal{N}}^{\infty} \rho (\alpha) \mathrm{d}\alpha &=& \int_{\phi_e}^{\infty}P(\phi,\mathcal{N})\mathrm{d}\phi  \label{eq:FPT_FP_int_equivalence} \\
\Rightarrow \rho (\mathcal{N}) &=& -\dfrac{\partial}{\partial \mathcal{N}} \int_{\phi_e}^{\infty}P(\Phi,\mathcal{N})\mathrm{d}\phi \label{eq:FPT_FP_pdf_equivalence}
\end{eqnarray}
To see why this is the case consider that the LHS of (\ref{eq:FPT_FP_int_equivalence}) is simply the probability that it takes longer than $\mathcal{N}$ e-folds for the field to reach $\phi_e$. If this field is the inflaton then this is simply the probability that inflation last longer than $\mathcal{N}$, $P($ inflation $ > \mathcal{N})$. The RHS is the area under the F-P PDF above the exit point $\phi_e$ at the time $\mathcal{N}$. Provided there is an absorbing boundary condition at $\phi_e$ then this area is the fraction of trajectories that have not yet reached $\phi_e$ at time $\mathcal{N}$. This means that all of these trajectories will take longer than $\mathcal{N}$ to reach $\phi_e$ so this area does indeed equal $P($ inflation $ > \mathcal{N})$ so the LHS = RHS. If there is not an absorbing boundary condition then the RHS will be larger than the LHS. This is because the RHS will now include contributions from trajectories that have reached $\phi_e$ previously but are now at $\phi > \phi_e$ meaning we are no longer computing a true \emph{first}-passage-time quantity. Therefore if one does not include an absorbing boundary condition, a computation of the RHS would \emph{overestimate} the number of trajectories that have yet to reach $\phi_e$ and the prediction for $\rho (\mathcal{N})$ would have a fatter tail than the true value.\\
\subsection{\label{sec:FPT linear}Stochastic processes with linear classical term}
Consider a general stochastic process of the form:
\begin{eqnarray}
    \dfrac{\mathrm{d}\tilde{\phi}}{\mathrm{d}\alpha} = -\gamma \lb \tilde{\phi} - \bar{\phi}_0\rb + A_{\tilde{\phi}}\xi \label{eq:gen linear stochastic}
\end{eqnarray}
where $\xi$ is a Gaussian white noise with unit variance. This stochastic process can be equivalently described in terms of the following F-P equation:
\begin{eqnarray}
\dfrac{\partial P}{\partial \alpha} = \vert \gamma \vert \dfrac{\partial}{\partial \chi} (\chi P)+ \dfrac{\vert \gamma \vert}{2}\dfrac{\partial^2 P}{\partial \chi^2}
\end{eqnarray}
where 
\begin{eqnarray}
\chi \equiv S(\gamma)\dfrac{\sqrt{\vert \gamma \vert}}{A_{\tilde{\phi}}}\lb \tilde{\phi} - \bar{\phi}_0 \rb
\end{eqnarray}
and $S(\gamma) = 1$ for $\gamma > 0$ and $S(\gamma) = -1$ for $\gamma < 0$.\\
If we perform the transformation 
\begin{eqnarray}
P (\chi,\alpha) = P_0\exp \lb \dfrac{\vert \gamma \vert}{2}\alpha - \dfrac{1}{2}\chi^2 \rb \Psi (\chi,\alpha) \label{eq:PHJ_transform}
\end{eqnarray}
where $P_0$ is some constant then $\Psi (\chi,\alpha)$ obeys:
\begin{eqnarray}
-\dfrac{1}{\vert \gamma \vert}\dfrac{\partial\Psi}{\partial\alpha} = \dfrac{1}{2}\lb -\dfrac{\partial^2}{\partial \chi^2} + \chi^2\rb \Psi
\end{eqnarray}
which reduces the problem to the quantum mechanical kernel for the simple harmonic oscillator. The free propagator, which for stochastic processes is known as the \textit{Mehler heat kernel} \cite{Pauli2000}, is given by:
\begin{eqnarray}
K(\chi,\alpha;\chi_{in},\alpha_{in}) &=& \dfrac{1}{\sqrt{2\pi A(\Delta \alpha)}}\exp \lb -\dfrac{1}{2}B(\Delta \alpha)\lb\chi^2 + \chi_{in}^2\rb + \dfrac{\chi\chi_{in}}{A(\Delta \alpha)}\rb \\
A(\Delta \alpha) &\equiv &\sinh \lb \vert \gamma \vert\Delta \alpha\rb \\
B(\Delta \alpha) &\equiv &\coth \lb \vert \gamma \vert\Delta \alpha\rb \\
\Delta \alpha &\equiv & \alpha - \alpha_{in}
\end{eqnarray}
However because we have an absorbing boundary at $\phi_e$, $\Psi$ is not given by the free kernel. To accommodate the absorbing boundary we use the method of images to add another free kernel mirrored so as to cancel at the boundary $\chi_e$:
\begin{eqnarray}
\Psi (\chi, \alpha) &=& K(\chi,\alpha;\chi_{in},\alpha_{in}) - K(2\chi_e -\chi,\alpha;\chi_{in},\alpha_{in}) \\
&=& \dfrac{1}{\sqrt{2\pi A(\Delta \alpha)}} \Bigg\lbrace \exp \lb -\dfrac{1}{2}B(\Delta \alpha)\lb\chi^2 + \chi_{in}^2\rb + \dfrac{\chi\chi_{in}}{A(\Delta \alpha)}\rb \nonumber \\
&&- \exp \lb -\dfrac{1}{2}B(\Delta \alpha)\lsb (2\chi_e -\chi)^2 + \chi_{in}^2\rsb + \dfrac{(2\chi_e -\chi)\chi_{in}}{A(\Delta \alpha)}\rb \Bigg\rbrace
\end{eqnarray}
We can then use (\ref{eq:PHJ_transform}) to obtain $P$:
\begin{eqnarray}
P (\chi,\alpha) &=& \dfrac{\sqrt{\vert \gamma \vert}}{A_{\tilde{\phi}}}\exp \lb \dfrac{\vert \gamma \vert}{2}\Delta \alpha - \dfrac{1}{2} (\chi^2 - \chi_{in}^2) \rb \Psi (\chi, \alpha )\\
&=&\dfrac{\sqrt{\vert \gamma \vert(n+1)}}{\sqrt{\pi} A_{\tilde{\phi}}}e^{-n\chi_{in}^{2}}\Bigg\lbrace \exp \lsb -(n+1)\chi^2 + 2(n+1)\chi \chi_{in}e^{-\vert \gamma \vert\Delta \alpha}\rsb \nonumber \\
&&-\exp \lsb -2(2n+1)\chi_{e}^{2} +4(n+1)\chi_e\chi_{in}e^{-\vert \gamma \vert\Delta \alpha}\rsb \nonumber \\
&& \times \exp \lsb  -(n+1)\chi^2 + \chi\lb 2(2n+1)\chi_e -2(n+1)\chi_{in}e^{-\vert \gamma \vert\Delta \alpha}\rb\rsb \Bigg\rbrace \label{eq:P_HJ}
\end{eqnarray}
where we were able to determine the constant $P_0 = \sqrt{\vert \gamma \vert}/A_{\tilde{\phi}}$ by use of the initial condition:
\begin{eqnarray}
P(\chi, \alpha \rightarrow \alpha_{in}) = P_0\delta (\chi - \chi_{in}) = P_0\dfrac{A_{\tilde{\phi}}}{\sqrt{\vert \gamma \vert}}\delta (\phi - \phi_{in})
\end{eqnarray}
We have also used the definitions of the hyperbolic trig functions so that we can rewrite $A$ and $B$ in terms of a new parameter $n$:
\begin{eqnarray}
n(\Delta \alpha) &\equiv &\dfrac{1}{e^{2\vert \gamma \vert\Delta {\alpha}}-1} \\
\dfrac{1}{A(\Delta \alpha)} &=& 2\lsb n(\Delta \alpha) + 1\rsb e^{-\vert \gamma \vert\Delta {\alpha}}\\
B(\Delta \alpha) &=& 2n(\Delta \alpha) + 1 
\end{eqnarray}
Equation (\ref{eq:P_HJ}) looks horribly complex and too difficult to integrate according to (\ref{eq:FPT_FP_pdf_equivalence}), fortunately however if you peer at it long enough you realise that it is actually just the sum of Gaussian integrals so that:
\begin{eqnarray}
\rho (\mathcal{N}) &=& -\dfrac{\partial}{\partial \mathcal{N} }\lcb \dfrac{1}{2}\text{erfc} \lsb\sqrt{n + 1}\bar{U} \rsb  - e^Y\dfrac{1}{2}\text{erfc} \lsb \sqrt{n} \bar{V}\rsb \rcb \\
Y &\equiv & -\dfrac{2n+1}{n+1}\chi_{e}^{2} -2n\chi_e \chi_{in}e^{-\vert \gamma \vert\Delta \mathcal{N}} \\
\bar{U} &\equiv & \chi_e - \chi_{in}e^{-\vert \gamma \vert\Delta \mathcal{N}}\\
\bar{V} &\equiv & \chi_{in} - \chi_e e^{-\vert \gamma \vert\Delta \mathcal{N}}
\end{eqnarray}
where $n = n(\Delta \mathcal{N})$. Evaluating the derivative we obtain:
\begin{eqnarray}
\rho (\mathcal{N}) &=& \dfrac{\vert \gamma \vert}{\sqrt{\pi}}\exp \lsb -(n+1)\bar{U}^2\rsb \lsb n\sqrt{n+1}\chi_e -\sqrt{n}(n+1)\chi_{in} \rsb \nonumber \\
&& -\dfrac{\vert \gamma \vert}{\sqrt{\pi}}\exp \lsb -n\bar{V}^2\rsb \lsb \sqrt{n}(n+1)\chi_{in}-n\sqrt{n+1}\lb 2-e^{-2\vert \gamma \vert\Delta \mathcal{N}}\rb\chi_e \rsb e^Y \nonumber \\
&& +\vert \gamma \vert\chi_e\lsb \chi_e e^{-2\vert \gamma \vert\Delta\mathcal{N}}-n(2n+3)\chi_{in}e^{-\vert \gamma \vert\Delta\mathcal{N}}\rsb e^Y \text{erfc}\lsb \sqrt{n}\bar{V}\rsb \label{eq:rhoN HJ_append}
\end{eqnarray}
Currently $\rho (\mathcal{N})$ is not normalised to 1 so we note that (\ref{eq:rhoN HJ_append}) should be divided by the quantity:
\begin{eqnarray}
    \lan 1\ran = 1 + \text{erf}\lb \chi_{e}^2\rb + e^{-\chi_{e}^2} \label{eq:norm_append}
\end{eqnarray}
although the maximum deviation from 1 for $\lan 1 \ran$ is $\sim 1.15$ at $\chi_e = 1/\sqrt{\pi}$ -- this is illustrated in Fig.~67 of \cite{Wilkins2023}. \\

To compute the abundance of PBHs from this FPT probability we simply use (\ref{eq:massfracdef}) to obtain:
\begin{eqnarray}
\beta &=& \dfrac{2\lsb e^{-\tilde{\Omega}^2\tilde{\mu}^2}-\text{erfc}(\tilde{\Omega}\tilde{\mu}) - e^{Y_{c}}\text{erfc}(\sqrt{n_c}\bar{V}_c) + \text{erfc}(\sqrt{n_c + 1}\bar{U}_c) \rsb}{1 + \text{erf}(\tilde{\Omega}\tilde{\mu}) + e^{-\tilde{\Omega}^2\tilde{\mu}^2}},\label{eq:massfrac_exactFULL_linear}\\
Y_c &\equiv & -\tilde{\Omega}^2\tilde{\mu}^2 \lsb 1 + e^{-2\vert \gamma \vert(\zeta_c + \left\langle \mathcal{N}\right\rangle)} +\dfrac{2n_c}{\tilde{\mu}} e^{-\vert \gamma \vert(\zeta_c + \left\langle \mathcal{N}\right\rangle)} \rsb, \label{eq: Y defn_linear} \\
\bar{U}_c &\equiv & \tilde{\Omega}\lsb \tilde{\mu} -  e^{-\vert \gamma \vert(\zeta_c + \left\langle \mathcal{N}\right\rangle)} \rsb,\label{eq: U defn_linear}\\
\bar{V}_c &\equiv & \tilde{\Omega}\lsb 1 - \tilde{\mu} e^{-\vert \gamma \vert(\zeta_c + \left\langle \mathcal{N}\right\rangle)} \rsb,\label{eq: V defn_linear} \\
n_{c} &\equiv &\dfrac{1}{e^{2\vert \gamma \vert(\zeta_c + \left\langle \mathcal{N}\right\rangle)} -1}, \label{eq:nc defn_linear} 
\end{eqnarray}
where $\zeta_c$ is the cutoff for PBH formation, $\lan \mathcal{N}\ran$ is the average number of e-folds to reach $\tilde{\phi}_e$ from $\tilde{\phi}_{in}$ and we have introduced the dimensionless parameters $\tilde{\Omega}$ and $\tilde{\mu}$:
\begin{eqnarray}
\tilde{\Omega} &\equiv  & S(\gamma)\dfrac{\sqrt{\vert \gamma \vert}}{A_{\tilde{\phi}}}\lb \tilde{\phi}_{in} - \bar{\phi}_0\rb = \dfrac{1}{\sqrt{\gamma}A_{\tilde{\phi}}}\Big\vert \dfrac{\mathrm{d}\tilde{\phi}}{\mathrm{d}\tilde{\alpha}}\Big \vert_{\alpha = \alpha_{in}}\label{eq: chi_in defn_linear} \\
\tilde{\mu} &\equiv &\dfrac{\tilde{\phi}_e - \bar{\phi}_0 }{\tilde{\phi}_{in} - \bar{\phi}_0} = e^{-\gamma N_{cl}} \label{eq:sigma defn_linear}
\end{eqnarray}
where $N_{cl}$ is the classical number of e-folds between $\tilde{\phi}_{in}$ and $\tilde{\phi}_{e}$. \\

During a phase of CR $\gamma  = \varepsilon_2/2$ and the initial field velocity is $\sqrt{2\varepsilon_1(\alpha_{in})}$. This is investigated more fully in section \ref{sec:CR}. The case of USR corresponds to the limit of $\gamma  = 3$ and substituting this into (\ref{eq:massfrac_exactFULL_linear}) recovers the results of \cite{Rigopoulos2021} and is given in section \ref{sec:PBH_nospec}.
\subsection{\label{sec:reallyCR}Stochastic processes with constant classical term}
Consider a general stochastic process of the form:
\begin{eqnarray}
    \dfrac{\mathrm{d}\tilde{\phi}}{\mathrm{d}\alpha} = - \gamma  + A_{\tilde{\phi}}\xi \label{eq:gen constant stochastic}
\end{eqnarray}
where $\xi$ is a Gaussian white noise with unit variance. This stochastic process can be equivalently described in terms of the following F-P equation:
\begin{eqnarray}
\dfrac{\partial P}{\partial \alpha} =  \dfrac{\partial P}{\partial \chi} + \dfrac{B}{2}\dfrac{\partial^2 P}{\partial \chi^2}
\end{eqnarray}
where 
\begin{eqnarray}
\chi &\equiv &\dfrac{1}{\gamma }\tilde{\phi} \\
B &\equiv & \dfrac{A_{\tilde{\phi}}^2}{\gamma^2}
\end{eqnarray}
If we perform the transformation 
\begin{eqnarray}
P (\chi,\alpha) = P_0\exp \lsb -\dfrac{1}{B}\lb \dfrac{\alpha}{2} + \chi \rb \rsb \Psi (\chi,\alpha) \label{eq:PHJ_transform_constant}
\end{eqnarray}
where $P_0$ is some constant then $\Psi (\chi,\alpha)$ obeys:
\begin{eqnarray}
\dfrac{\partial\Psi}{\partial\alpha} = \dfrac{B}{2}\dfrac{\partial^2\Psi}{\partial \chi^2} 
\end{eqnarray}
which has a fundamental solution given by the free heat kernel:
\begin{eqnarray}
K(\chi,\alpha;\chi_{in},\alpha_{in}) = \dfrac{1}{\sqrt{2\pi B\Delta \alpha}}\exp \lsb -\dfrac{\lb \chi - \chi_{in}\rb^2}{2B\Delta \alpha}\rsb 
\end{eqnarray}
where $\Delta \alpha = \alpha - \alpha_{in}$.
To obtain $\Psi (\chi,\alpha)$ we must impose an absorbing boundary at $\chi_e$:
\begin{eqnarray}
\Psi (\chi, \alpha) &=& K(\chi,\alpha;\chi_{in},\alpha_{in}) - K(2\chi_e -\chi,\alpha;\chi_{in},\alpha_{in}) \\
\Rightarrow \Psi (\chi, \alpha) &=& \dfrac{1}{\sqrt{2\pi B\Delta \alpha}}\lcb \exp \lsb -\dfrac{\lb \chi - \chi_{in}\rb^2}{2B\Delta \alpha}\rsb - \exp \lsb -\dfrac{\lb 2\chi_e - \chi - \chi_{in}\rb^2}{2B\Delta \alpha}\rsb  \rcb \label{eq:Psi_const}
\end{eqnarray}
Utilising equations (\ref{eq:FPT_FP_pdf_equivalence}), (\ref{eq:PHJ_transform_constant}) \& (\ref{eq:Psi_const}), we can compute the first-passage time to reach $\tilde{\phi}_e$ like so:
\begin{eqnarray}
    \rho (\mathcal{N}) &=& -\dfrac{\partial}{\partial \mathcal{N}}\lcb \dfrac{P_0}{2}e^{-\mathcal{N}/B} \lsb \text{erfc} ~T_{+}(\mathcal{N}) - e^{-2\chi_e/B}\lb 2 - \text{erfc}~ T_{-}(\mathcal{N}) \rb\rsb\rcb \\
   \Rightarrow  \rho (\mathcal{N})  &=& \dfrac{P_0}{2}e^{-\mathcal{N}/B}\Bigg\lbrace \dfrac{1}{B}\lsb \text{erfc} ~T_{+}(\mathcal{N}) - e^{-2\chi_e/B}\lb 2 - \text{erfc}~ T_{-}(\mathcal{N}) \rb\rsb \nonumber \\
    & & \quad \quad \quad \quad \quad + \dfrac{2}{\sqrt{\pi}}\lsb e^{-T_{+}^2(\mathcal{N})}\partial_{\mathcal{N}}T_{+}(\mathcal{N}) - e^{-2\chi_e/B}e^{-T_{-}^2(\mathcal{N})}\partial_{\mathcal{N}}T_{-}(\mathcal{N})\rsb \Bigg\rbrace 
    \end{eqnarray}
    where
    \begin{eqnarray}
    {T}_{\pm}(\mathcal{N}) &=& \dfrac{\sqrt{\mathcal{N}}}{\sqrt{2B}}\lsb\dfrac{\chi_e -\chi_{in}}{\mathcal{N}} \pm 1\rsb \\
   \Rightarrow \partial_{\mathcal{N}}T_{\pm}(\mathcal{N}) &=& \dfrac{\sqrt{\mathcal{N}}}{2\sqrt{2B}}\lsb\dfrac{\chi_e -\chi_{in}}{\mathcal{N}}\lb 1 - \dfrac{2}{\mathcal{N}}\rb  \pm 1\rsb\\
\dfrac{P_0}{2} &=&  \dfrac{1}{1-\exp \lb -2\chi_e /B\rb}
\end{eqnarray}
Which tells us that the abundance of PBHs can be given by:
\begin{eqnarray}
    \beta_{\scalebox{0.5}{$\varepsilon_2 = 0$}} &=&  \dfrac{e^{-\mathcal{N}_c/B}}{1-\exp \lb -2\chi_e /B\rb} \lcb \text{erfc} ~T_{+}(\mathcal{N}_c) - e^{-2\chi_e/B}\lsb 2 - \text{erfc}~ T_{-}(\mathcal{N}_c) \rsb\rcb \\
    \mathcal{N}_c &=& \zeta_c + \lan\mathcal{N}\ran \approx \zeta_c + \chi_{in} - \chi_e 
\end{eqnarray}
where the approximation utilises the fact that $\mathcal{N}$ is very close to the classical number of e-folds. 
\section{Sudden end approximation}
\label{app:sudden-end}

The sudden-end approximation requires that the phase transition of the waterfall field happens extremely quickly after $M^2(\phi,\sigma)$ becomes negative. By looking at the classical\footnote{The stochastic effects are only important at the instant $M^2 = 0$ to give an initial displacement to $\chi$. Before and after this instant the motion is dominated by the classical terms.} equation of motion for the waterfall field: 
\begin{eqnarray}
    0=\dfrac{\mathrm{d}^2\chi}{\mathrm{d}\alpha^2}+3\dfrac{\mathrm{d}\chi}{\mathrm{d}\alpha}+M^2(\phi,\sigma)\chi+\lambda\chi^3,\label{eq:class_chi_eom}\\ 
\end{eqnarray}
we can see that this requires $M^2(\phi,\sigma)$ to become large (and negative) quickly after the field value $\phi_e$ is reached. This will allow us to ignore the dynamics of the waterfall field in our calculations. To investigate the value of $M^2(\phi,\sigma)$ a short time $\delta \alpha$ after $M^2(\phi_e,\sigma_e) = 0$ we refer to the equations of motion for the inflaton and the spectator field:
\begin{eqnarray}    
    \dfrac{\mathrm{d}\phi}{\mathrm{d}\alpha} &=& -\dfrac{\varepsilon_2}{2}(\phi-\phi_0) + A_{\phi}\xi_1,\label{eq:app_phi_eom}\\    
    \dfrac{\mathrm{d}\sigma}{\mathrm{d}\alpha} &=& -\dfrac{1}{2}\lb 3 - 2\nu \rb \sigma + A_{\sigma}\xi_2. \label{eq:app_sigma_eom}
\end{eqnarray}
and the definition of $M^2$:
\begin{eqnarray}
    M^2 = -m^{2}_{\chi} + \lambda_{\phi}\phi^2 -\lambda_{\sigma}\sigma^2 \label{eq:app_M_defn}
\end{eqnarray}
From (\ref{eq:app_M_defn}) it is clear that the value of $M^2(\phi,\sigma)$ a small time $\delta \alpha$ after $M^2(\phi_e,\sigma_e) = 0$ is given by the formula:
\begin{eqnarray}
    \delta \lb M^2\rb = 2\lambda_{\phi}\lsb \phi_e \delta \phi + \lb\delta \phi \rb^2 - \tilde{\lambda}\sigma_e\delta \sigma  -\tilde{\lambda}\lb\delta \sigma \rb^2 \rsb \label{eq:app_delM1}
\end{eqnarray}
where it is necessary to expand to second order to obtain $\mathcal{O}(\delta \alpha)$ terms as equations (\ref{eq:app_phi_eom}) \& (\ref{eq:app_sigma_eom}) have stochastic terms in. This can be straightforwardly done:
\begin{eqnarray}
    \delta \phi &=& \lsb -\dfrac{\varepsilon_2}{2}(\phi-\phi_0) + A_{\phi}\xi_1\rsb \delta \alpha + \mathcal{O}(\delta \alpha^2) \\
    (\delta \phi )^2 &=& A_{\phi}^2 \delta \alpha + \mathcal{O}(\delta \alpha^2) \\
    \delta \sigma &=& \lsb -\dfrac{1}{2}\lb 3 - 2\nu \rb \sigma + A_{\sigma}\xi_2\rsb \delta \alpha + \mathcal{O}(\delta \alpha^2) \\
    (\delta \sigma )^2&=& A_{\sigma}^2 \delta \alpha + \mathcal{O}(\delta \alpha^2) 
\end{eqnarray}
These equations can be substituted into (\ref{eq:app_delM1}) and taking the expectation value of this we obtain\footnote{Assuming $\varepsilon_2 > 0$.}:
\begin{eqnarray}
    \lan \delta \lb M^2\rb \ran = -\lambda_{\phi}A_{\phi}^2\lsb \Omega\mu \dfrac{m_{\chi}}{\sqrt{\lambda_{\phi}}}\dfrac{\sqrt{2\varepsilon_2}}{A_{\phi}} -2 + R\lb \dfrac{\varepsilon_2}{3-2\nu}-3\rb \rsb\delta \alpha + \mathcal{O}(\delta \alpha^2) \label{eq:app_delM_soln}
\end{eqnarray}
where $\Omega$ and $\mu$ are defined in equations (\ref{eq: Omega defn CR}) \& (\ref{eq: mu defn CR}) such that $\Omega\mu \sim \mathcal{O}(1)$ for PBH production and $R \leq 1$ is defined in (\ref{eq: R defn}). It is easy to convince oneself that the terms in the square bracket combine to make a large and positive value -- mainly due to the large value of the first term\footnote{Recall that $A_{\phi} \leq 10^{-5}$ while $\Omega\mu$ cannot be smaller than $\mathcal{O}(1)$ to prevent overproduction of PBHs.} -- and therefore we expect the sudden-end approximation to still be valid. After one e-fold $\delta\alpha=1$, $\chi$ will have settled in its minimum if $\left|M^2(\phi,\sigma)\right| \gg 3H^2$ which we can see is indeed the case:
\begin{eqnarray}
    \dfrac{\left\vert\lan \delta \lb M^2\rb \ran \right \vert}{3H^2} \simeq \dfrac{\lambda_{\phi}}{12\pi^2}\lsb \Omega\mu \dfrac{m_{\chi}}{\sqrt{\lambda_{\phi}}}\dfrac{\sqrt{2\varepsilon_2}}{A_{\phi}} -2 + R\lb \dfrac{\varepsilon_2}{3-2\nu}-3\rb \rsb \gg 1\label{eq:exp_sol_M}
\end{eqnarray}
Without a spectator the condition for the sudden-end approximation to be valid is simply $m_{\chi}^2/\sqrt{\lambda_{\phi}} \gg 2/A_{\phi}$. The additional spectator term in (\ref{eq:exp_sol_M}) will usually \emph{accelerate} the transition provided $\varepsilon_2/3 > 3 - 2\nu$. This condition is satisfied for $\varepsilon_2 \geq 1.5$ when $\nu = 5/4$ which means only the $\varepsilon_2 = 1$ case studied in this paper does not accelerate the transition. However for $\varepsilon_2 = 1$ the third term in (\ref{eq:app_delM_soln}) becomes $-R$ and as $R\leq 1$ this does not significantly alter the transition rate.


\bibliographystyle{ieeetr85}
\bibliography{Bibliography}
\end{document}